\documentclass[a4]{article}
\usepackage[square,numbers]{natbib}
\usepackage[dvips]{graphicx}
\usepackage{psfrag}
\usepackage{pstricks}
\usepackage{epsfig}
\usepackage{amsmath}
\usepackage{amssymb}
\usepackage{latexsym}
\usepackage{mathtools}
\usepackage{overpic}
\usepackage{stmaryrd}
\usepackage{authblk}

\usepackage{url}
\usepackage{color}

\title{Patch-recovery filters for curvature in discontinuous Galerkin-based level-set methods}

\author[1,2,+]{F. Kummer\thanks{The autor acknowledges financial support by the German Research Foundation (DFG).}}
\author[2]{T. Warburton}
\affil[1]{
Department of Mechanical Engineering,
Technische Universit{\"a}t Darmstadt,
Otto-Berndt-Stra{\ss}e 2, 64287 Darmstadt, Germany
}
\affil[2]{
Department of Computational and Applied Mathematics,
Rice University,
Houston, TX 77005-1892, U.S.A.
}
\affil[+]{
(corresponding author, kummer@fdy.tu-darmstadt.de)
}

\newcommand{\real}{\mathbb{R}}
\newcommand{\frakA}{\mathfrak{A}}
\newcommand{\frakB}{\mathfrak{B}}
\newcommand{\frakI}{\mathfrak{I}}

\newcommand{\GridCells}{{\mathfrak{K}_h}}

\newcommand{\charfunc}{{\boldsymbol{1}}}
\newcommand{\jump}[1]{\left\llbracket {#1} \right\rrbracket}
\newcommand{\mean}[1]{\left\{\!\left\{{#1}\right\}\!\right\}}

\newcommand{\dx}{\ \mathrm{d} \vec{x}}
\newcommand{\Kref}{K_{\mathrm{ref}}}
\renewcommand{\vec}[1]{\boldsymbol{#1}}
\newcommand{\nI}{{\vec{n}_{\frakI}}}
\newcommand{\divergence}[1]{{\mathrm{div}\left({#1}\right)}}

\newcommand{\XDG}[1]{{\mathbb{P}^{X}_{#1}}}
\newcommand{\DGP}[1]{{\mathbb{P}_{#1}}}
\newcommand{\DGQ}[1]{{\mathbb{Q}_{#1}}}
\newcommand{\SEM}[1]{{\mathbb{Q}^{\mathcal{C}^0}_{#1}}}

\newcommand{\dS}{{ \ \mathrm{dS} }}
\newcommand{\dl}{{ \ \mathrm{d} \ell}}
\newcommand{\dV}{{ \ \mathrm{dV} }}

\newcommand{\normI}{{\vec{n}_{\frakI}}}

\newcommand{\normGamma}{{\vec{n}_{\Gamma}}}
\newcommand{\normGammaI}{{\vec{n}_{\frakI,\Gamma}}}

\newcommand{\patchrec}[1]{{{\mathrm{prc}}^{#1}}}
\newcommand{\Cnull}{{\mathcal{C}^0}}
\newcommand{\proj}[2]{{\mathrm{Proj}_{{ #1 }} \left( { #2 } \right) }}

\def\Xint#1{\mathchoice
{\XXint\displaystyle\textstyle{#1}}%
{\XXint\textstyle\scriptstyle{#1}}%
{\XXint\scriptstyle\scriptscriptstyle{#1}}%
{\XXint\scriptscriptstyle\scriptscriptstyle{#1}}%
\!\int}
\def\XXint#1#2#3{{\setbox0=\hbox{$#1{#2#3}{\int}$}
\vcenter{\hbox{$#2#3$}}\kern-.5\wd0}}
\def\dashint{\Xint\setminus}

\newcommand{\GradientOption}{$\vec{g} = ?$}
\newcommand{\LevSet}{$\nabla f$}
\newcommand{\FiltLevSet}{$\nabla \tilde{f}$}
\newcommand{\FiltLevSetH}{$\partial^2 \tilde{f}$}
\newcommand{\LevSetH}{${\partial^2 f}$}
\newcommand{\HessianOption}{$\vec{H} = ?$}
\newcommand{\FiltLevSetGrad} {{$\nabla \tilde{\vec{g}}$}}
\newcommand{\LevSetGrad} {{$\nabla {\vec{g}}$}}
\newcommand{\useFiltLevSetGrad}{fil. $\nabla$?}
\newcommand{\true}{true}
\newcommand{\false}{false}
\newcommand{\useFiltLevSetHess}{fil. $\partial^2$?}

\newcommand{\LevelSetSource}{$f = ?$}
\newcommand{\fromDG}{$\varphi_{\mathrm{br}}$}
\newcommand{\fromCnull}{$\varphi_{\mathcal{C}^0}$}

\DeclareSymbolFont{extraup}{U}{zavm}{m}{n}
\DeclareMathSymbol{\varheart}{\mathalpha}{extraup}{86}
\DeclareMathSymbol{\vardiamond}{\mathalpha}{extraup}{87}

\begin{document}

\maketitle

\begin{abstract}
In two-phase flow simulations, a difficult issue is usually the
treatment of surface tension effects.
These cause
a pressure jump that is proportional to the curvature of the interface separating the two fluids.
Since the evaluation of the curvature incorporates second derivatives,
it is prone to numerical instabilities.
Within this work, the interface is described by a level-set method based on
a discontinuous Galerkin discretization.
In order to stabilize the evaluation of the curvature,
a patch-recovery operation is employed.
There are numerous ways in which this filtering operation can be applied in
the whole process of curvature computation.
Therefore,
an extensive numerical study is performed to identify optimal settings for the patch-recovery operations
with respect to computational cost and accuracy.
\end{abstract}

\section{Introduction and motivating example}

In order to simulate immiscible two-phase flows, one usually has to
consider surface tension effects.
These induce a pressure jump
which is proportional to the total curvature $\kappa$ of the fluid interface $\frakI$.
Precisely, the momentum equation
for the fluid domains $\frakA$ and $\frakB$ is given as
\begin{equation}
  \frac{\partial}{\partial t} ( \rho \vec{u})
+ \divergence{\rho \vec{u} \otimes \vec{u} }
+ \nabla \psi
= \mu \Delta \vec{u},
\label{eq:Momentum}
\end{equation}
while at the fluid interface $\frakI = \overline{\frakA} \cap \overline{\frakB}$
the velocity $\vec{u}$ and pressure $\psi$ are coupled via the
jump condition
\begin{equation}
  \jump{
    \left( \psi \vec{I}
    -
    \mu \left( \nabla \vec{u} + \nabla \vec{u}^T \right)
    \right) \nI
  }
  =
  \sigma \kappa \nI,
\label{eq:momJump}
\end{equation}
see e.g. \cite{hutter_continuum_2004} or \cite{wang_thermodynamic_2011}.
We briefly introduce the notation required for this formula:
\begin{itemize}
  \item The computational domain $\Omega \subset \real^D$ is decomposed into
        fluid $\frakA$, fluid $\frakB$ and the $(D-1)$-dimensional interface $\frakI$,
        i.e. $\Omega = \frakA \cup \frakI \cup \frakB$.
        Regarding this work, we restrict ourselves to the case $D=2$.

  \item $\nI$ denotes the normal vector on $\frakI$, oriented so that
        ``it points from $\frakA$ to $\frakB$'' and $\kappa$ denotes the
        total curvature of $\frakI$.
        We assume $\frakI$ to be smooth enough, so that both,
        $\nI$ and $\kappa$ are at least in $\Cnull(\frakI)$.
  \item the jump operator for $u \in \mathcal{C}^0(\Omega \setminus \frakI)$ is defined as
\begin{equation}
        \jump{u}(\vec{x})
        =
        \lim_{\xi \searrow 0} \left(
        u(\vec{x} + \xi \nI) - u(\vec{x} - \xi \nI)
        \right).
\end{equation}
  \item $\mu$ and $\rho$ denote the dynamic viscosity and the density of the fluid, which are
        usually constant within either $\frakA$ and $\frakB$, but have a jump at the interface.
\end{itemize}

This setting may be described by a level-set function
$\varphi \in \mathcal{C}^2(\Omega)$, which fulfills
\begin{equation}
  \left\{ \begin{array}{rcl}
  \varphi < 0 & \mathrm{in} & \frakA \\
  \varphi > 0 & \mathrm{in} & \frakB \\
  \varphi = 0 & \mathrm{on} & \frakI \\
  \end{array}  \right.
\end{equation}
Then, obviously,
\begin{equation}
   \nI = \frac{\nabla \varphi }{ | \nabla \varphi |_2}
   \qquad \textrm{and} \qquad
   \kappa = \divergence{\nI}.
\label{eq:Bonnet}
\end{equation}
The latter expression is also called Bonnet's formula. In dependence of
the level-set gradient $\nabla \varphi$ and
the level-set Hessian $\partial^2 \varphi$,
it can be expressed as
\begin{equation}
   \divergence{ \frac{\nabla \varphi}{ | \nabla \varphi |_2} }
   =:
   \mathrm{curv}(\nabla \varphi, \partial^2 \varphi)
\end{equation}
with
\begin{equation}
  \mathrm{curv}(\vec{g},\vec{H}) = \frac{\mathrm{tr}(\vec{H})}{ | \vec{g} |_2}
              - \frac{\vec{g}^T \vec{H} \vec{g}}{ | \vec{g} |_2^3}.
\end{equation}
Note that by the introduction of
$\varphi$, the properties $\nI$ and $\kappa$, which were initially only defined on $\frakI$,
were smoothly extended to the whole domain $\Omega$.

The purpose of this paper is to benchmark various filtering strategies for Bonnet's
formula, based on patch recovery.
In order to assess the quality of a filtering strategy,
the $L^2$-error in the curvature may only be one property of interest;
in addition,
we define the following benchmark problem:
\begin{equation}
   \left\{ \begin{array} {rcll}
       \Delta \psi                    & =  & 0              & \textrm{in } \Omega \setminus \frakI \\
       \jump{ \psi }                  & =  & \sigma \kappa  & \textrm{on } \frakI                  \\
       \jump{ \nabla \psi \cdot \nI } & =  & 0              & \textrm{on } \frakI                  \\
       \psi                           & =  & 0              & \textrm{on } \partial \Omega         \\
   \end{array} \right.
   \label{eq:Poisson}
\end{equation}
This Poisson problem shares some behavior with a two-phase flow problem (\ref{eq:Momentum}),
since it
would correspond e.g. to the pressure-projection step
of a projection method.
It is solved by an extended discontinuous Galerkin method (XDG), details are given in section \ref{sec:XDGPoisson}.

\subsection{Outline for this work}
\label{sec:outline}

The main motivation for this work is given in section \ref{sec:motivation_example},
where it is demonstrated, for a simple example,
that different choices for the level-set-field $\varphi$, with equal zero-sets $\frakI$,
can lead to very different results if just a straightforward projection of the
curvature is used.
Next, the patch-recovery operation which we employ to overcome this issues
is introduced (section \ref{sec:patch_recovery}) and the test setup is defined (section \ref{sec:test_setup}).
Results and a discussion, together with recommendations for the use of patch-recover are given in section
\ref{sec:results_and_discussion}.
Finally, in appendix \ref{sec:XDGPoisson} the discretization of the Poisson problem (\ref{eq:Poisson})
is given.

\subsection{Treatment of surface tension in numerical methods.}
\label{sec:litrature_review}


In multi-phase flow simulations, surface tension plays
an important role as soon as length scales become small enough, e.g.
in the range of centimeters and below.
Over the previous decade, a huge variety of two-phase flow solvers have been
proposed and a significant portion of them uses level-set methods.
Some of them use a smeared interface approach
-- where the piecewice constant properties  $\rho$ and $\mu$ in
eqs. (\ref{eq:Momentum},\ref{eq:momJump})
are regularized so that their derivatives are finite --
like e.g. the works of \citet{herrmann_simulating_2011}.
In contrast to this, extended finite element or discontinuous Galerkin methods
use an approximation space that is conformal with the location of the  discontinuities,
like e.g. done by \citet{SvenGross06}.
Common to all this methods is that surface tension forces have to be computed from
a level-set field $\varphi$.

From a flow-solver perspective, what finally matters is the force which is induced into
the momentum balance (\ref{eq:Momentum}) by surface tension effects.
This force is only active at the interface $\frakI$ itself, i.e.
in some finite volume (FV), finite element (FE) or discontinuous Galerkin (DG) discretizations,
it usually appears at the right-hand-side of the momentum equation as
\[
  \oint_\frakI \sigma \kappa \normI \cdot \vec{v} \dS
   \quad  \textrm{ or }  \quad
  \int_\Omega \sigma \kappa \normI \cdot \vec{v}  \delta_{ \frakI } \dV
   .
\]
Here, $\vec{v}$ denotes a test function
(in 2D FV-methods usually, $\vec{v}=(1,0)$ and $\vec{v} = (0,1)$)
while $\delta_{ \frakI }$ denotes a delta-distribution on $\frakI$.
It is either singular,
thereby both integrals given above are equivalent, or regularized, i.e.
`smeared out' in a small neighborhood around $\frakI$.
Whether a `sharp' or a `smeared' approach is used usually
depends on the design of the flow solver, i.e.
whether the flow solver represents the density and viscosity jump in
a `sharp' or smeared fashion.

However, the evaluation of curvature is not the only way to compute surface tension forces.
Using a variant of Stoke's integral theorem,
one can re-express surface tension forces in some control volume $K \subset \Omega$ as
\begin{equation}
\oint_{\frakI \cap K} \sigma \kappa \normI \cdot \vec{v} \dS
=
\oint_{\frakI \cap K} \sigma (I - \normI \otimes \normI) : \nabla \vec{v} \dS
-
\dashint_{\partial (\frakI \cap K)} \sigma \vec{s} \cdot \vec{v} \dl,
\label{eq:SurfaceStokes}
\end{equation}
where $\vec{s}$ is tangential to $\frakI$ and normal onto ${\partial (\frakI \cap K)}$.
and $\dashint ... \dl$ denotes an integral over the boundary of the
surface $ \frakI \cap K$; i.e. in 2D, $\dashint ... \dl$ it is a 0-dimensional point integral (counting measure) over
the two end-points of the line $\frakI \cap K$, while in 3D it is the 1-dimensional
line integral over the boundary of the surface $\frakI \cap K$.
Since $\kappa \normI$ can also be expressed by
the Laplace-Beltrami - operator on the manifold $\frakI$
-- for details, we refer to  \citet{SvenGross06} --
the reformulation of the surface tension is often referred to as
the Laplace-Beltrami approximation of the surface tension terms.
The main benefit of this re-formulation is
that the right-hand-side of eq. (\ref{eq:SurfaceStokes}) does not depend on second
derivatives of $\varphi$.
This seems very attractive e.g. in a FV-discretiztion, where the test function is constant
and therefore the first term
on the right-hand-side of (\ref{eq:SurfaceStokes}) vanishes and the problem is reduced
to the computation of tangents $\vec{s}$ to $\frakI$.

The Laplace-Beltrami formulation also shows good results
in the extended finite element (XFEM) context,
as demonstrated e.g. by
\citet{SvenGross06},
\citet{cheng_xfem_2012},
\citet{sauerland_stable_2013}.

However, if the numerical integration over the surface $\frakI$ becomes more precise
-- as is the case in this work --
equation (\ref{eq:SurfaceStokes}) is also fulfilled numerically
and therefore one cannot expect significant `filtering properties'
from the Laplace-Beltrami formulation, i.e.
any oscillations present in the level-set
would be captured equally by both, the left- and the right-hand-side
of equation (\ref{eq:SurfaceStokes}).

In the work of \citet{chen_projection_2004}, a
projection
of the curvature
on the broken piecewise polynomial space
is obtained from the relation
\begin{equation}
\int_{K} \underbrace{ \divergence{ \frac{\nabla \varphi}{| \nabla \varphi |_2} } }_{= \kappa} v \dV
=
\oint_{\partial K}  \frac{\nabla \varphi}{| \nabla \varphi |_2 } \cdot \vec{n}_K v \dS
-
\int_{K}  \frac{\nabla \varphi}{| \nabla \varphi |_2 } \cdot \nabla v \dV
\label{eq:bonnetRed}
\end{equation}
for a test function $v$ and a control volume $K$ with outer normal $\vec{n}_K$.
Since this is equivalent to the
piecewise polynomial reconstruction
of
Bonnet's formula (\ref{eq:Bonnet}), one can expect that it is prone to the same issues as we
outline in section \ref{sec:motivation_example}.
However, the right-hand-side of equation \ref{eq:bonnetRed} provides the option
to use a non-local flux-formulation for $\frac{\nabla \varphi}{| \nabla \varphi |_2 }$
at the boundary integral.
This was exploited e.g. by
\citet{heimannThesis2013}, were an additional diffusion term is incorporated into
the non-local projection to deal with discontinuities.

A completely different approach to curvature evaluation  is proposed by
\citet{heimann_unfitted_2013}. They use an
extended DG space
 -- they refer to it as `unfitted DG' --
for velocity and pressure, a conventional DG field for the level-set
while for curvature computation, they employ finite difference scheme.

The patch-recovery post-processing was first introduced for finite element methods by
\citet{zienkiewicz_superconvergent_1992}.
To our knowledge, the only application of
patch recovery (see section \ref{sec:patch_recovery}) to curvature evaluation
has been demonstrated by \citet{JibbenHerrmann2012}.
There, a nodal projection is used, while this work employs an $L^2$-projection,
making the algorithm independent of any choice of nodes and more general
with respect to the shape of the patches.
The main achievement of this work is that we
also investigate the application of patch-recovery for `intermediate' values, like gradients
$\nabla \varphi$ and Hessians $\partial^2 \varphi$
of the level-set function $\varphi$.

\subsection{Piecewise polynomial approximation}
\label{sec:polynomial_spaces}
For sake of completeness, we briefly introduce the numerical grid
as well as the approximation spaces used e.g. for $\varphi$ or $\psi$.
These are fairly standard, and can be found in similar form in
many textbooks, see e.g. \citet{hesthaven_nodal_2008} or \citet{di_pietro_mathematical_2011}.
We define:
\begin{itemize}
\item
the numerical grid: $\GridCells = \left\{ K_1, \ldots, K_J \right\}$, with $h$ being the maximum
diameter of all cells $K_j$. The cells cover the whole domain ($\overline{\Omega} = \bigcup_j \overline{K_j}$),
but do not overlap ($\int_{K_j \cap K_l} 1 \dx = 0$ for $l \neq j$).
We restrict ourselves to non-curved grids, i.e.,
each cell $K_j$ can be described as the image of a reference cell
$\Kref$ under an affine-linear mapping $T_j : \real^2 \rightarrow \real^2$,
i.e. $T_j(\Kref) = K_j$;

\item the broken piecewise polynomial
approximation space of maximum total degree $p$:
\begin{equation}
  \DGP{p}(\GridCells) := \left\{
    f \in L^2(\Omega); \
    \forall \ K \in \GridCells:
    f|_{K} = \sum_{0 \leq i + j \leq p} x^{i} y^{j} b_{i j},
    \ b_{i j} \in \real
  \right\}
\end{equation}

\item the broken piecewise polynomial approximation
 space of maximum degree $p$:
\begin{equation}
  \DGQ{p}(\GridCells) := \left\{
    f \in L^2(\Omega); \
    \forall \ K \in \GridCells:
    f|_{K} = \sum_{0 \leq i, j \leq p} x^{i} y^{j} b_{i j},
    \  b_{i j} \in \real
  \right\}
\end{equation}

\item the continuous piecewise polynomial
space of order $p$:
\begin{equation}
 \SEM{p}(\GridCells) := \DGQ{p}(\GridCells) \cap \Cnull(\Omega)
\end{equation}

\item the extended broken approximation
 space of order $p$, which is used to discretize
the Poisson problem (\ref{eq:Poisson}):
\begin{equation}
 \XDG{p}(\varphi,\GridCells) := \DGP{p}(\GridCells) \charfunc_{\frakA} \oplus \DGP{p}(\GridCells) \charfunc_{\frakB},
\end{equation}
where $\charfunc_X$  denotes the characteristic function for set $X \subset \Omega$.

\item the projector onto some subspace $X$ of $L^2(\Omega)$ in the $L^2$-norm:
\begin{equation}
   L^2(\Omega) \ni f \mapsto \proj{X}{f} \in X.
\end{equation}

\item
   that all derivative operators ($\divergence{-}$, $\nabla$, $\Delta$ and the Hessian $\partial^2$)
   in this work should be understood in a broken sense.
   We assert that all properties defined here are almost everywhere sufficiently smooth,
   so that the respective derivatives exist in a classical sense everywhere up to a set of measure zero.
   Therefore, we do not make the usual distinctions, e.g. between the gradient $\nabla$
   and the broken gradient $\nabla_h$.
\end{itemize}

\subsection{Motivating example.}
\label{sec:motivation_example}
The main issue with using Bonnet's formula is that the results are extremely sensitive
to minor disturbances in the level-set function $\varphi$, which we are going to illustrate
by the following example.
The exact interface $\frakI_{\mathrm{ex}}$ is chosen to be a circle with radius $R=0.8$ around the
origin. Obviously, there are infinitely many choices for $\varphi$.

For a sharp-interface method,
like the one described in appendix \ref{sec:XDGPoisson},
the interface $\frakI$ has to be a closed surface,
therefore it is required that also the numerical representation
of the level-set field is at least continuous, i.e. $\varphi \in \Cnull(\Omega)$.
To ensure this, the exact level-set function $\varphi_{\mathrm{ex}}$ is projected
onto the continuous piecewise polynomial space $\SEM{2}(\GridCells)$.

Two representations of a circular interface  around the origin with radius $8/10$ are compared,
namely:
\[
\begin{array}{lrll}
   \textrm{case (a)}:     & \varphi = &                             (8/10)^2 - x^2 - y^2
       & \textrm{quadratic}, \ \varphi = \varphi_{\mathrm{ex}}       \\
   \textrm{case (b)}:     & \varphi = & \proj{\SEM{2}(\GridCells)}{8/10 - \sqrt{x^2 - y^2}}
       & \textrm{signed-distance}, \ \varphi \neq \varphi_{\mathrm{ex}}\\
\end{array}
\]
The domain $\Omega = (-3/2, 3/2)^2$ is discretized by $18 \times 18$ equidistant
quadrilateral cells.
Obviously, the exact curvature at the exact interface
in this example $\kappa|_{\frakI} = -10/8$, and the exact solution to
the Poisson problem (\ref{eq:Poisson}), for $\sigma = 1/10$ is
\begin{equation}
   \psi_{\mathrm{ex}} =
   \left\{ \begin{array}{ll}
      0     & \textrm{in } \frakA \\
      1/8   & \textrm{in } \frakB \\
   \end{array} \right. .
\end{equation}

Note that in case (a), the quadratic form already is a
member of the space $\SEM{2}(\GridCells)$, therefore the circular interface is represented
\emph{exactly} in the polynomial approximation space.
In case (b),
where this is not the case,
 the approximation of the circular interface still seems reasonably good:
for any $\vec{x}$ in the zero-set of $\varphi$, the
error in the radius, i.e. $| \ |\vec{x}|_2 - 8/10 \ |$ is less than $1.2 \cdot 10^{-4}$.

\begin{figure}
\begin{centering}
\input{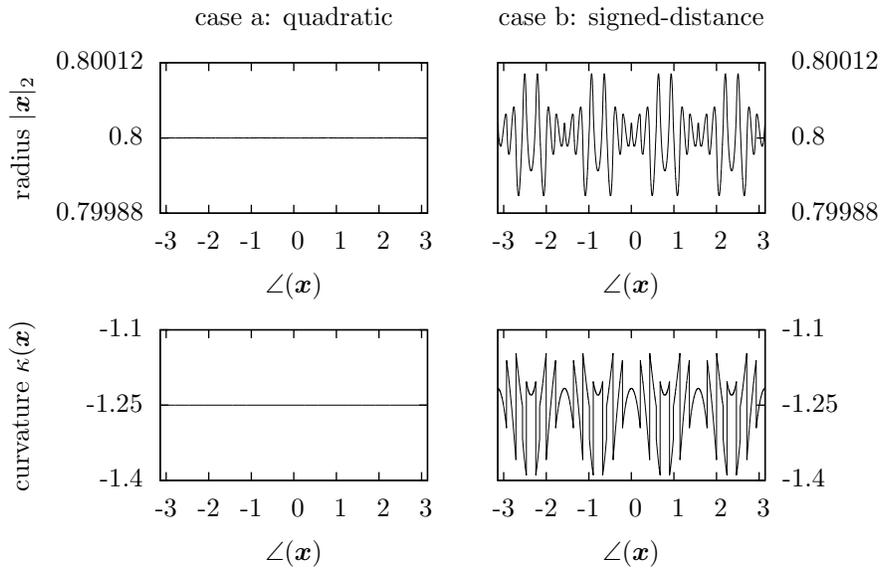}
\end{centering}
\caption{
Polar plot of a quadratic versus a signed-distance level-set function $\varphi$,
for the circular interface with radius $0.8$:
for points $\vec{x} \in \frakI$, the radius $| \vec{x} |_2$ and the curvature at $\vec{x}$,
i.e. $\kappa(\vec{x})$ are plotted in dependence of the angular coordinate $\angle(\vec{x})$
For the quadratic representation, the interface is exact
up to round-off errors, yielding a quite accurate representation of the curvature
in a piecewise polynomial space with sufficiently high order.
On the other hand, for the signed distance representation, an error
with a magnitude of about $10^{-4}$ in the radius yields an $L^\infty$-error
around 10 \% in the curvature.
}
\label{fig:example1}
\end{figure}

\begin{figure}
\begin{centering}
\[
\begin{array}{lcc}
&  \textrm{case (a): quadratic} & \textrm{case (b): signed-distance} \\
&  \varphi = (8/10)^2 - x^2 - y^2 &
     \varphi = \proj{\SEM{2}(\GridCells)}{8/10 - \sqrt{x^2 + y^2}}  \\
\rotatebox{90}{\rule{1.1cm}{0.0cm} level-set field $\varphi$} &
\begin{overpic}[width=0.49\textwidth
]{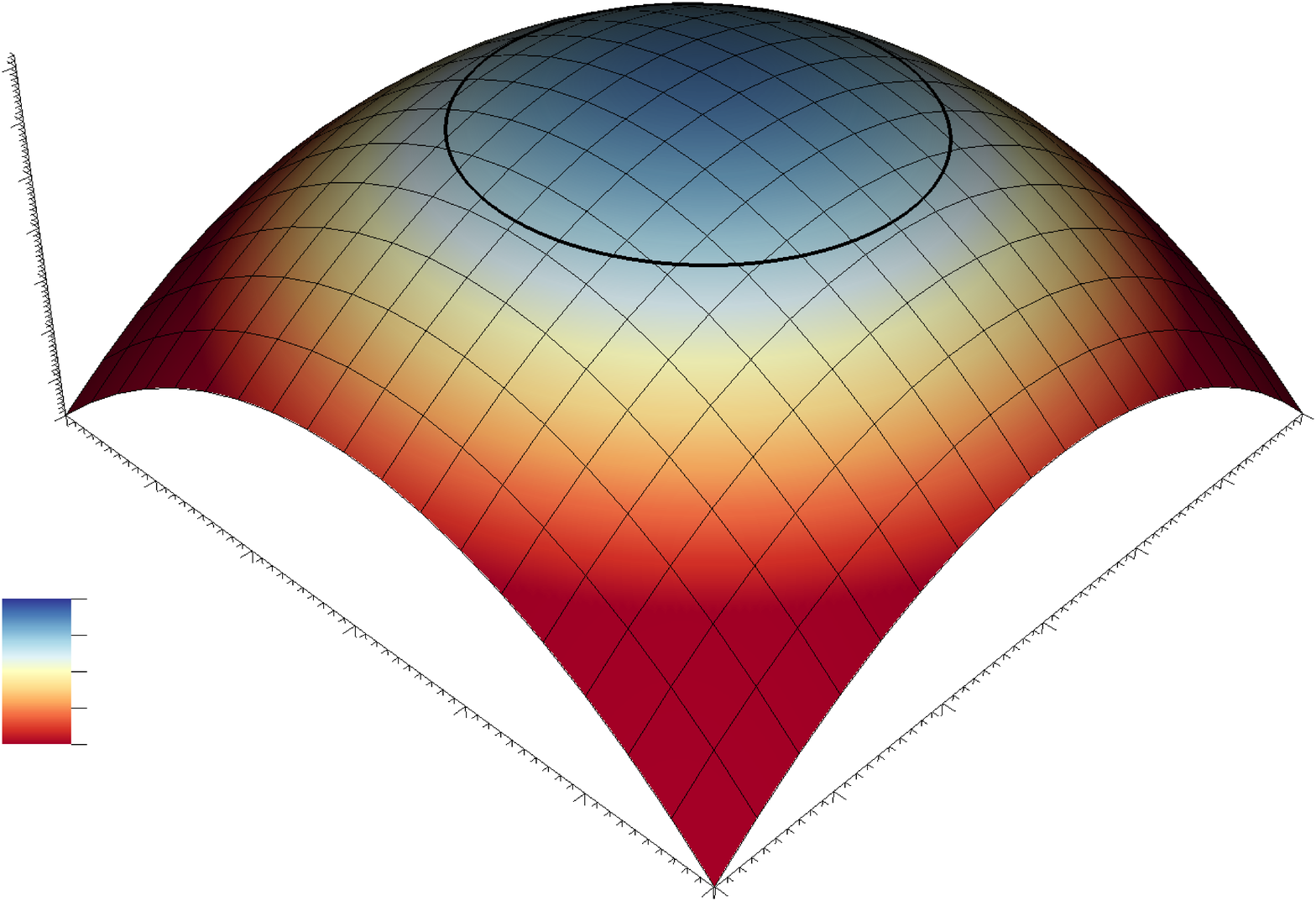}
\put(7,22)  {\scriptsize $1$}
\put(7,16)  {\scriptsize $0.5$}
\put(7,10.8){\scriptsize $-2$}
\put(-2.5,33)  {\scriptsize $1.5$}
\put(24,17)    {\scriptsize $0$}
\put(44,-3)    {\scriptsize $-1.5$}
\put(99,33)    {\scriptsize $1.5$}
\put(79,17)    {\scriptsize $0$}
\put(55,-1)    {\scriptsize $-1.5$}
\put(-2.5,57)  {\scriptsize $0$}
\put(-6.5,38)  {\scriptsize $-3.3$}
\put(2,62)     {\scriptsize $\varphi$}
\end{overpic} & \rule{0.3cm}{0cm}
\begin{overpic}[width=0.49\textwidth
]{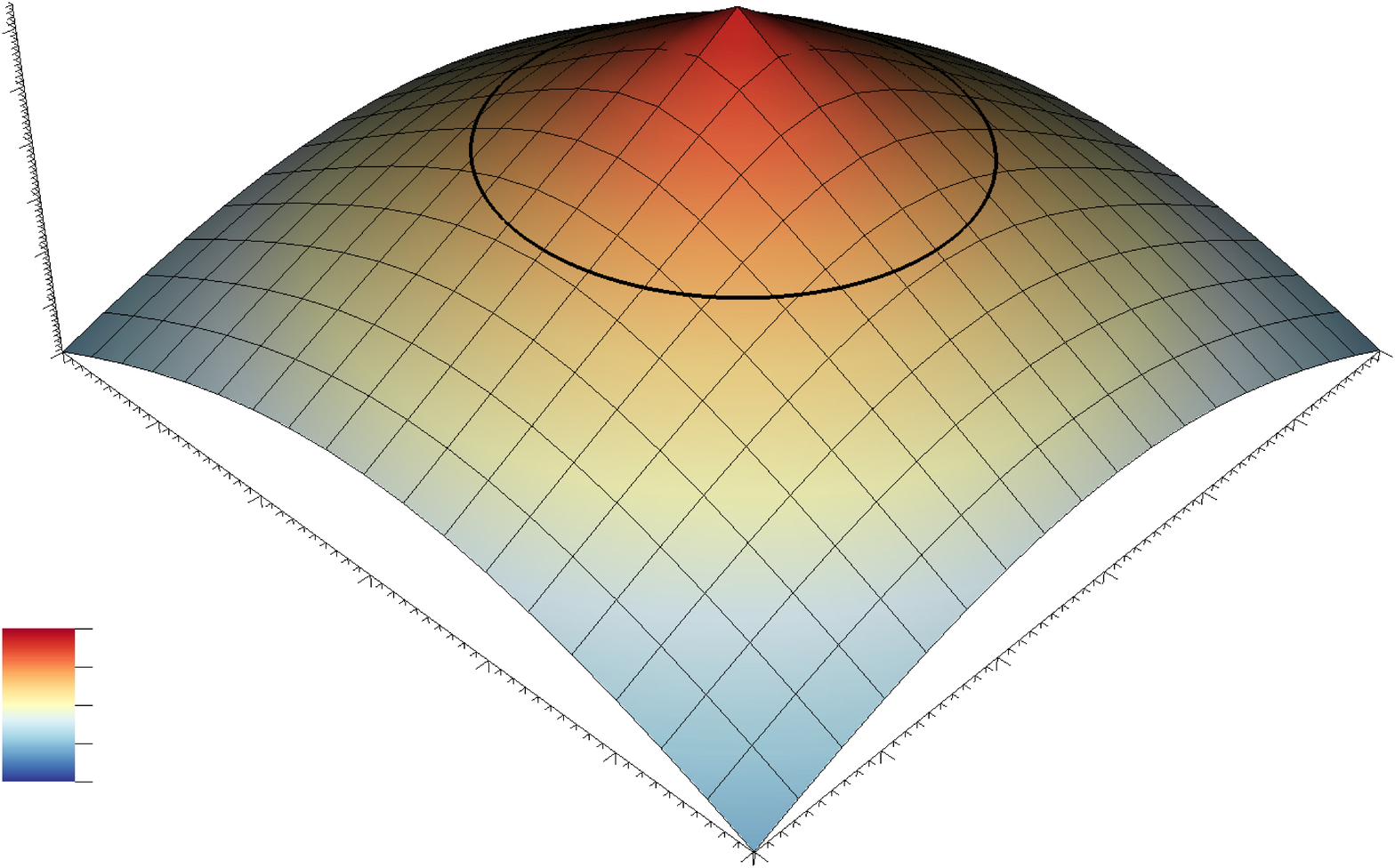}
\put(7,16)  {\scriptsize $1$}
\put(7,10)  {\scriptsize $0.5$}
\put(7,4.8) {\scriptsize $-2$}
\put(-2.5,33)  {\scriptsize $1.5$}
\put(24,17)    {\scriptsize $0$}
\put(44,-3)    {\scriptsize $-1.5$}
\put(99,33)    {\scriptsize $1.5$}
\put(79,17)    {\scriptsize $0$}
\put(55,-1)    {\scriptsize $-1.5$}
\put(-6,58)    {\scriptsize $0.8$}
\put(-1.5,50)  {\scriptsize $0$}
\put(-6.5,39)  {\scriptsize $-1.2$}
\put(2,62)     {\scriptsize $\varphi$}
\end{overpic}
\\ 
 & & \\
\rotatebox{90}{\rule{1.2cm}{0.0cm} curvature $\kappa$} &
\begin{overpic}[width=0.49\textwidth
]{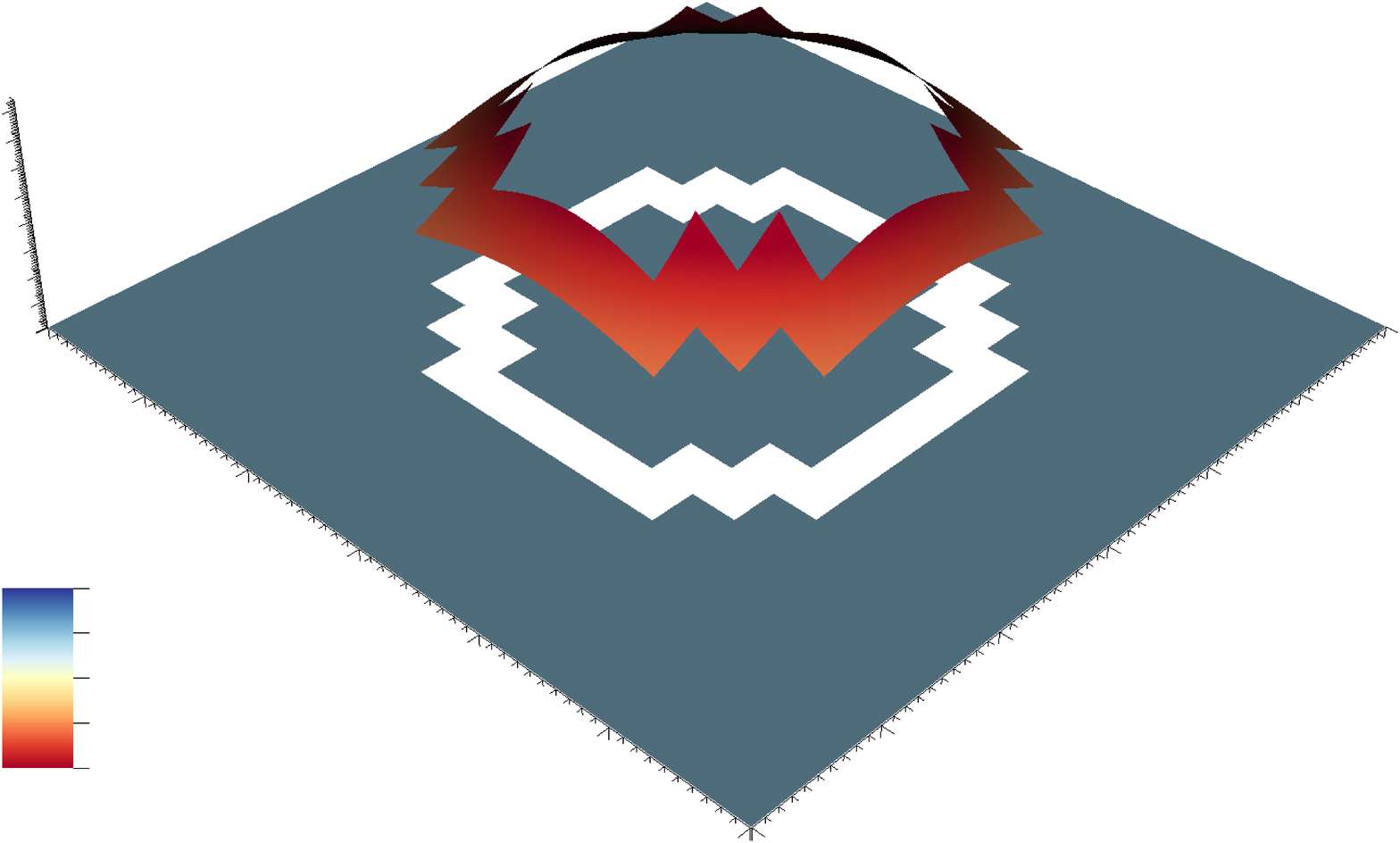}
\put(7,17)  {\scriptsize $0.5$}
\put(7,11)  {\scriptsize $-0.5$}
\put(7,4.5) {\scriptsize $-1.5$}
\put(-2.5,33)  {\scriptsize $1.5$}
\put(24,17)    {\scriptsize $0$}
\put(44,-3)    {\scriptsize $-1.5$}
\put(99,33)    {\scriptsize $1.5$}
\put(79,17)    {\scriptsize $0$}
\put(55,-1)    {\scriptsize $-1.5$}
\put(55,-1)    {\scriptsize $-1.5$}
\put(-9,47)    {\scriptsize $-1.2$}
\put(-8.5,36)  {\scriptsize $-0.2$}
\put(2,50)     {\scriptsize $\kappa$}
\end{overpic} & \rule{0.3cm}{0cm}
\begin{overpic}[width=0.49\textwidth
]{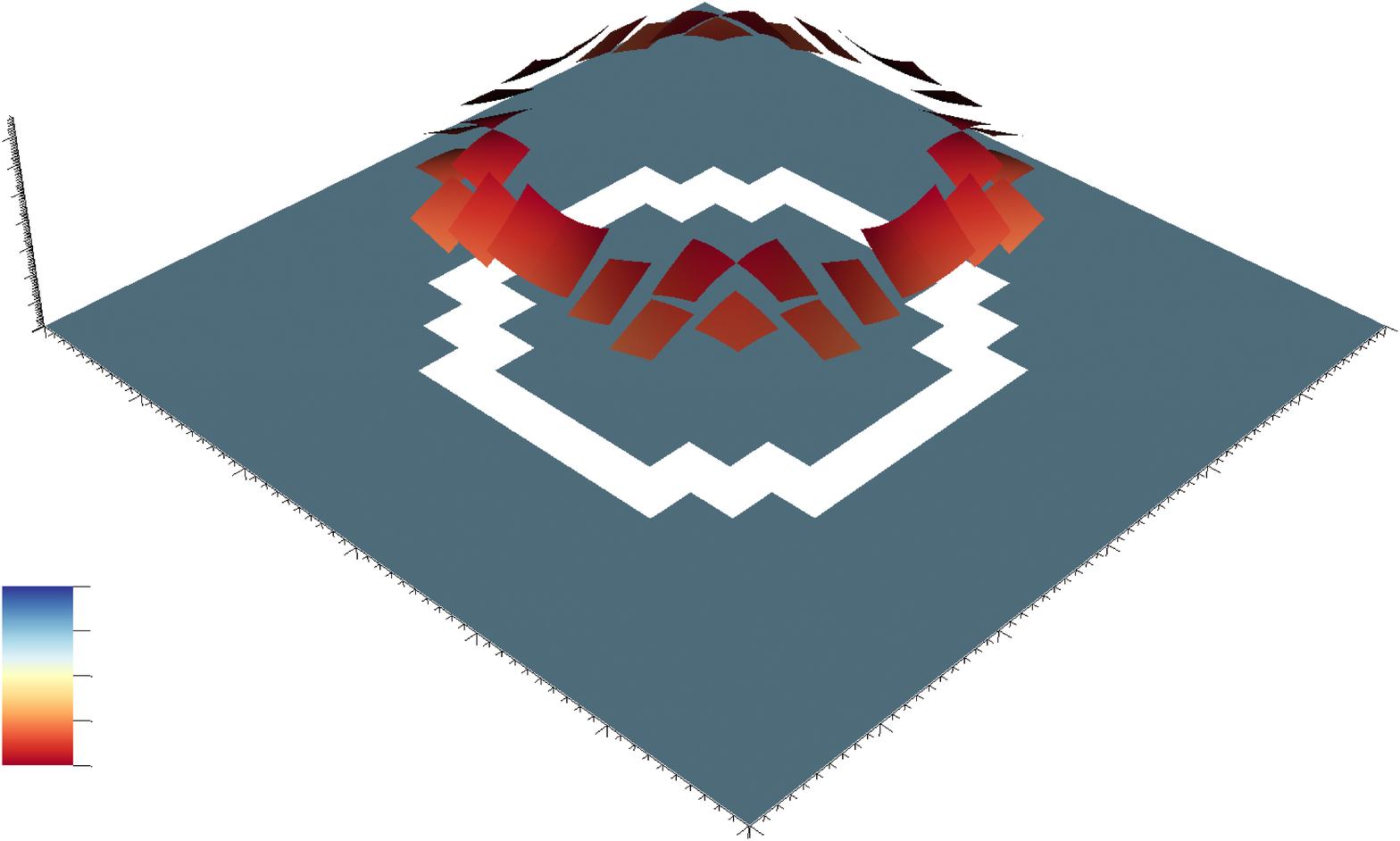}
\put(7,17)  {\scriptsize $0.5$}
\put(7,11){\scriptsize $-0.5$}
\put(7,4.5){\scriptsize $-1.5$}
\put(-2.5,33)  {\scriptsize $1.5$}
\put(24,17)    {\scriptsize $0$}
\put(44,-3)    {\scriptsize $-1.5$}
\put(99,33)    {\scriptsize $1.5$}
\put(79,17)    {\scriptsize $0$}
\put(55,-1)    {\scriptsize $-1.5$}
\put(-9,47)    {\scriptsize $-1.2$}
\put(-8.5,36)  {\scriptsize $-0.2$}
\put(2,50)     {\scriptsize $\kappa$}
\end{overpic}
\\ 
 & & \\
\rotatebox{90}{\rule{1.2cm}{0.0cm} pressure $\psi$} &
\begin{overpic}[width=0.49\textwidth
]{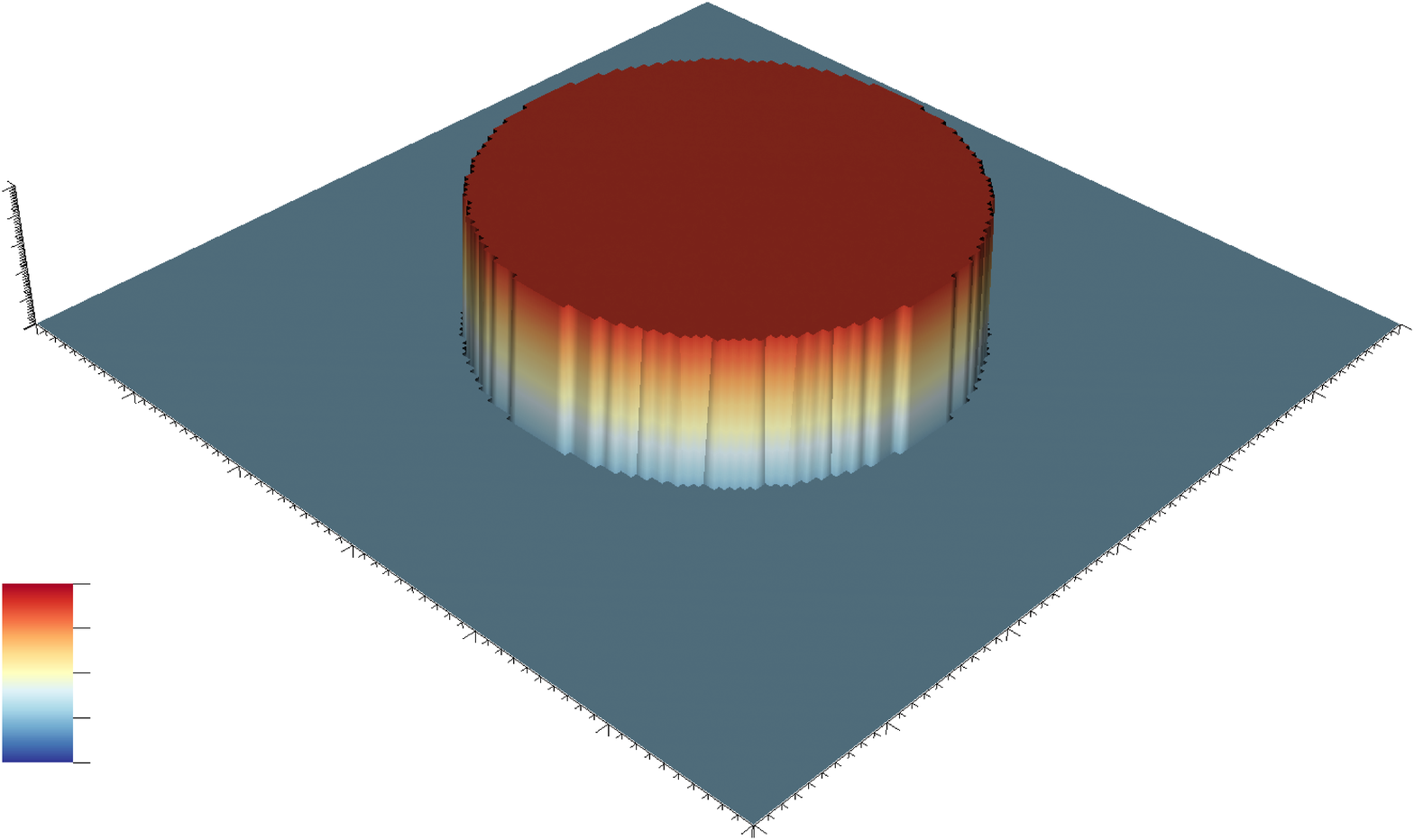}
\put(7,17)  {\scriptsize $0.15$}
\put(7,10.5){\scriptsize $0.05$}
\put(7,4.5) {\scriptsize $-0.05$}
\put(-2.5,33)  {\scriptsize $1.5$}
\put(24,17)    {\scriptsize $0$}
\put(44,-3)    {\scriptsize $-1.5$}
\put(99,33)    {\scriptsize $1.5$}
\put(79,17)    {\scriptsize $0$}
\put(55,-1)    {\scriptsize $-1.5$}
\put(-6,42.5)  {\scriptsize $0.1$}
\put(2,45)     {\scriptsize $\psi$}
\end{overpic} & \rule{0.3cm}{0cm}
\begin{overpic}[width=0.49\textwidth
]{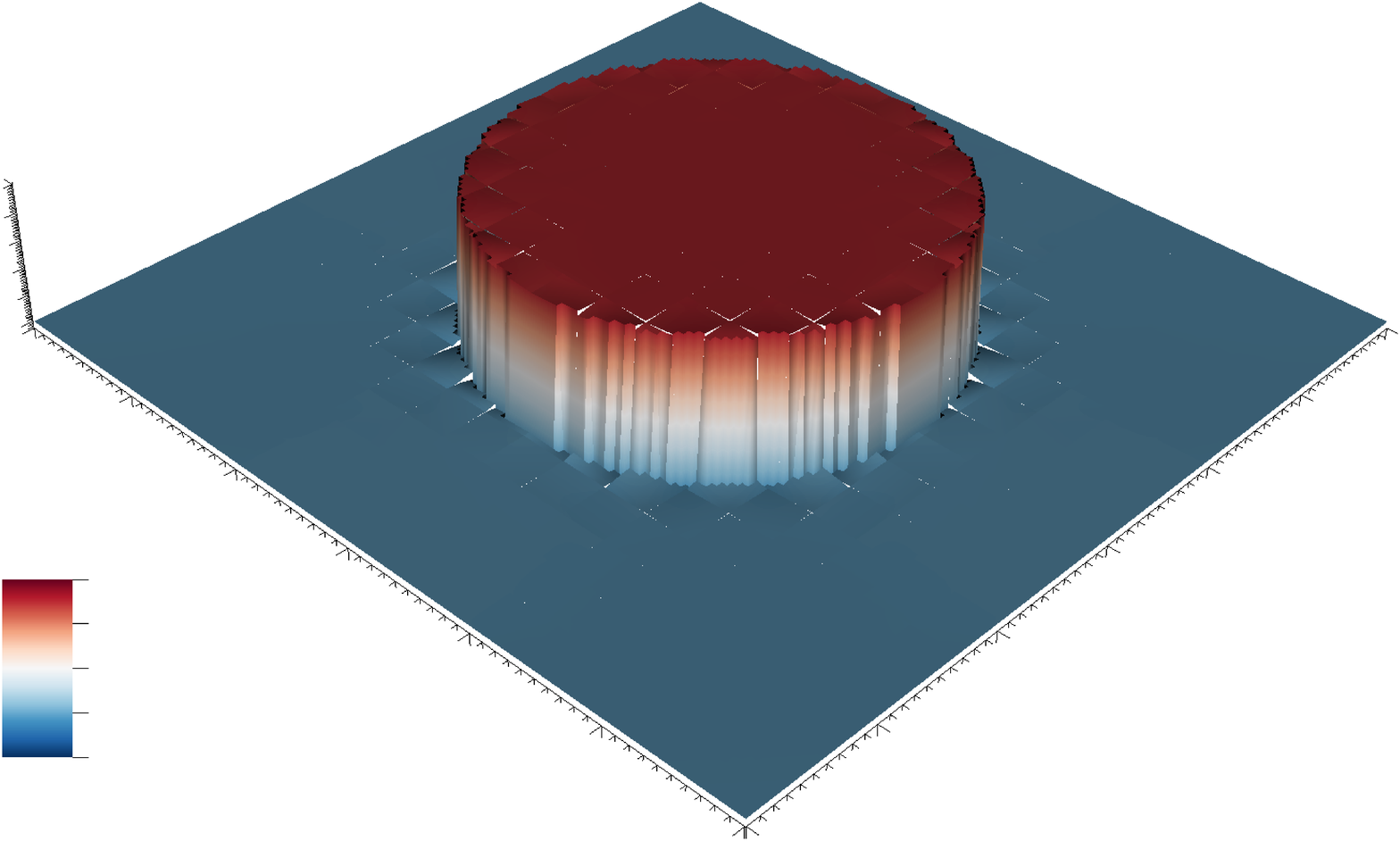}
\put(7,17)  {\scriptsize $0.15$}
\put(7,10.5){\scriptsize $0.05$}
\put(7,4.5) {\scriptsize $-0.05$}
\put(-2.5,33)  {\scriptsize $1.5$}
\put(24,17)    {\scriptsize $0$}
\put(44,-3)    {\scriptsize $-1.5$}
\put(99,33)    {\scriptsize $1.5$}
\put(79,17)    {\scriptsize $0$}
\put(55,-1)    {\scriptsize $-1.5$}
\put(-6,42.5)  {\scriptsize $0.1$}
\put(2,45)     {\scriptsize $\psi$}
\end{overpic}
\\ 
 & & \\
\rotatebox{90}{\rule{0.5cm}{0.0cm} pressure gradient $\frac{\partial \psi}{\partial x}$} &
\begin{overpic}[width=0.49\textwidth
]{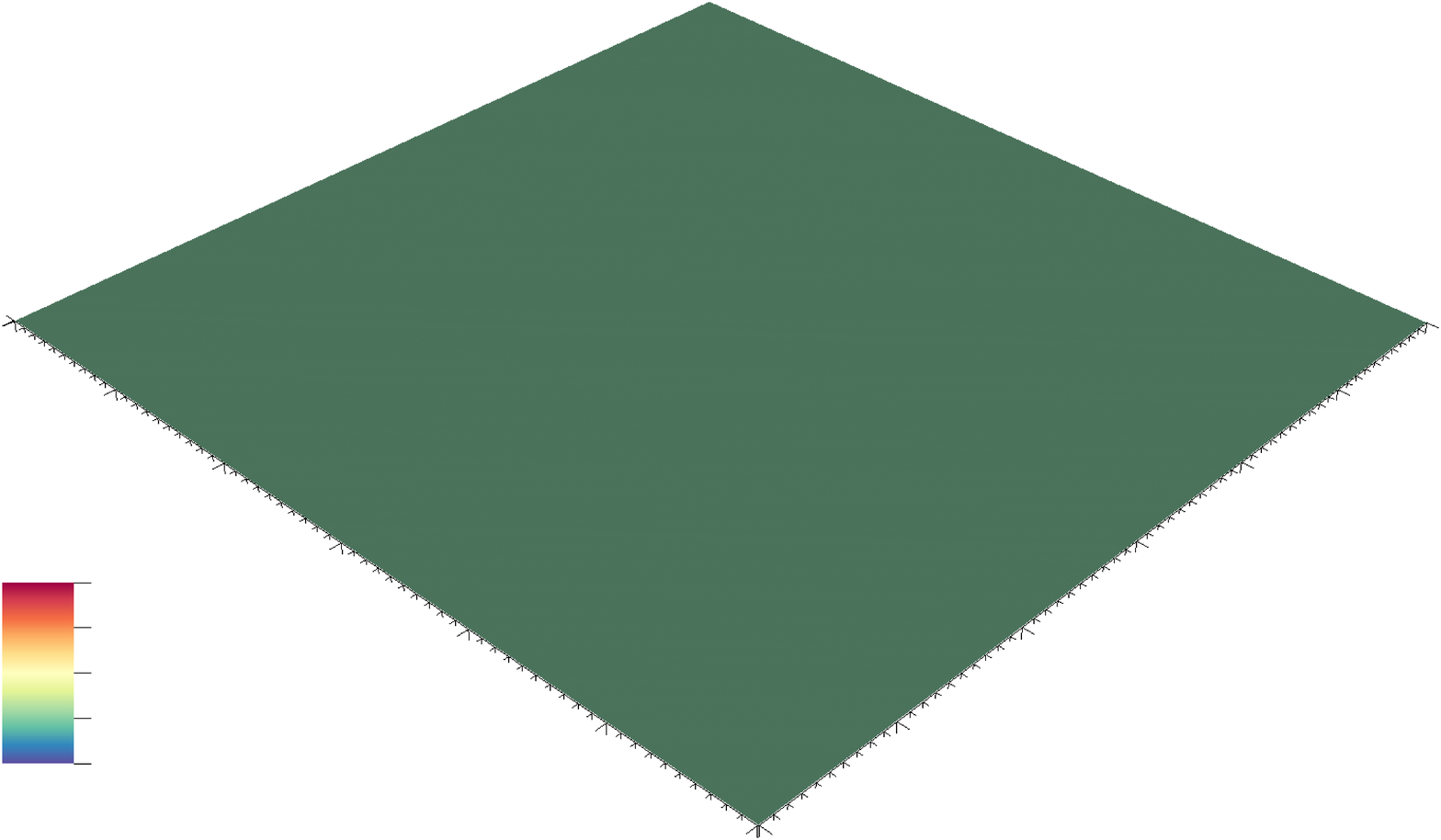}
\put(7,17)  {\scriptsize $0.4$}
\put(7,10.5){\scriptsize $0$}
\put(7,4.5) {\scriptsize $-0.4$}
\put(-2.5,33)  {\scriptsize $1.5$}
\put(24,17)    {\scriptsize $0$}
\put(44,-3)    {\scriptsize $-1.5$}
\put(99,33)    {\scriptsize $1.5$}
\put(79,17)    {\scriptsize $0$}
\put(55,-1)    {\scriptsize $-1.5$}
\put(2,50)     {\scriptsize $\frac{\partial \psi}{\partial x}$}
\end{overpic} & \rule{0.3cm}{0cm}
\begin{overpic}[width=0.49\textwidth
]{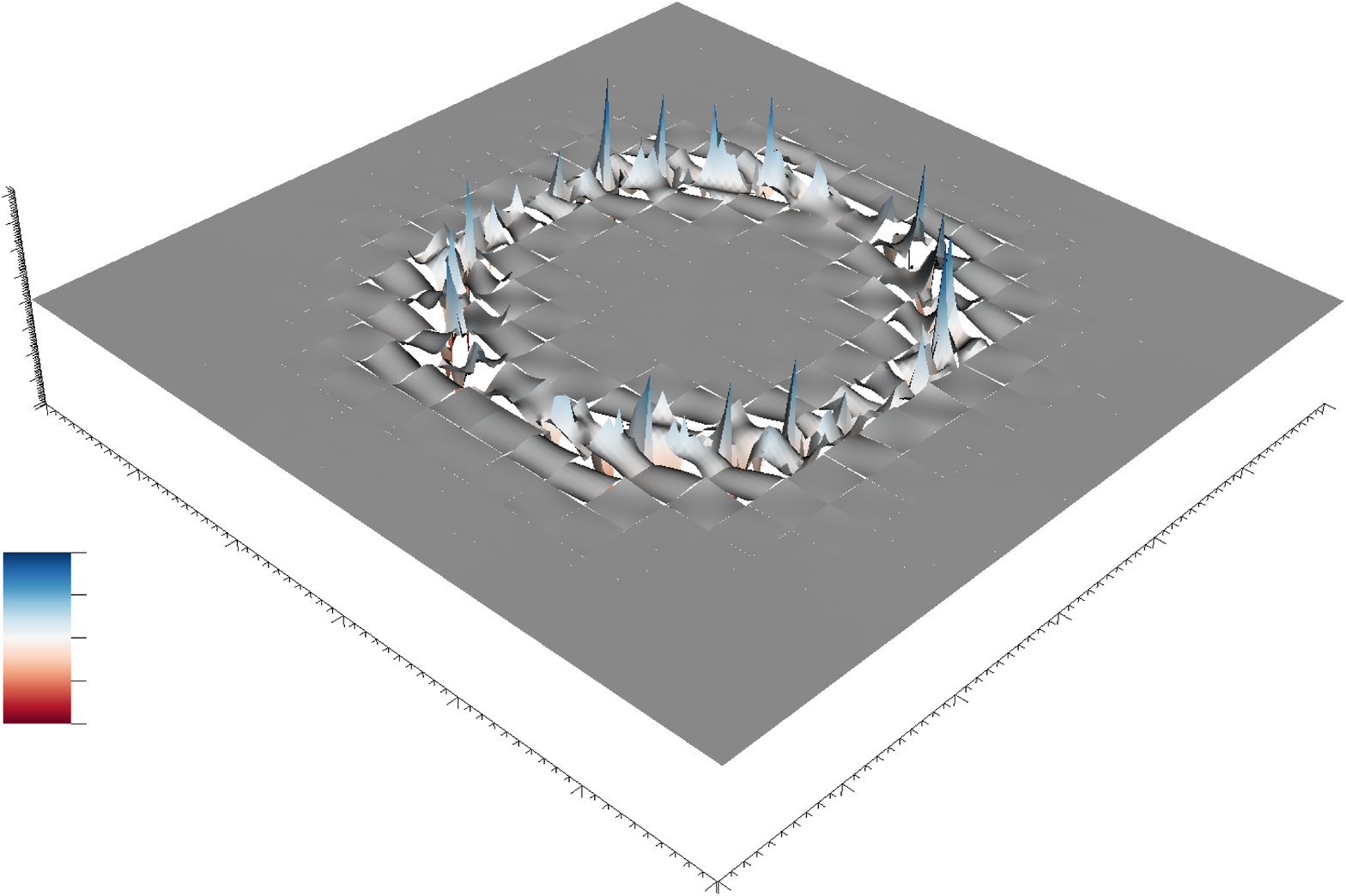}
\put(7,24)  {\scriptsize $0.4$}
\put(7,18)  {\scriptsize $0$}
\put(7,11.5){\scriptsize $-0.4$}
\put(-2.5,33)  {\scriptsize $1.5$}
\put(24,17)    {\scriptsize $0$}
\put(44,-3)    {\scriptsize $-1.5$}
\put(99,33)    {\scriptsize $1.5$}
\put(79,17)    {\scriptsize $0$}
\put(55,-1)    {\scriptsize $-1.5$}
\put(55,-1)    {\scriptsize $-1.5$}
\put(-6,48)    {\scriptsize $0.3$}
\put(-8.5,36.5){\scriptsize $-0.3$}
\put(2,50)     {\scriptsize $\frac{\partial \psi}{\partial x}$}
\end{overpic}
\\
\end{array}
\]
\end{centering}
\caption{
Comparison of a quadratic versus a signed-distance level-set function $\varphi$.
While the zero level-set of both representation is pretty similar,
the error in the curvature $\kappa$ for case (b) is obvious.
In succession, this causes high oscillations in the pressure gradient
of the Poisson problem whose right-hand-side depends on $\kappa$.
}
\label{fig:example1_b}
\end{figure}

However, if in sub-sequence the curvature
\begin{equation}
\kappa := \proj{
 \DGP{12}(\GridCells)
} {
\mathrm{curv} ( \nabla \varphi ,  \partial^2 \varphi )
}
\end{equation}
is computed, and the Poisson problem (\ref{eq:Poisson})
with a right-hand-side based on $\kappa$ is solved,
a quite reasonable error in $\psi$, and, more severe, $\nabla \psi$ is introduced,
which can also be seen in figures \ref{fig:example1} and \ref{fig:example1_b}:
\begin{center}
\begin{tabular}{ll|c|c}
error type                 &
                           & quadratic                & signed-distance             \\
  \hline
radius              & $\forall \vec{x} \in \frakI: \ | \ |\vec{x}|_2 - 8/10 \  | $
                           & $\leq 10^{-11}$          & $\leq 1.2 \cdot 10^{-4}$    \\
curvature           & $\forall \vec{x} \in \frakI: \ \left| \ |\kappa(\vec{x})| - (-10/8) \ \right|$
                           & $\leq 2.2 \cdot 10^{-9}$ & $\leq 0.14$                 \\
pressure            & $\left\| \psi - \psi_{\mathrm{ex}} \right\|_{L^2(\Omega)}$
                           & $\leq 10^{-10}$          & $\approx 1.3 \cdot 10^{-3}$ \\
pressure gradient   & $\left\| \nabla \psi - \nabla \psi_{\mathrm{ex}} \right\|_{\infty}$
                           & $\leq 2 \cdot 10^{-8}$   & $\approx 0.4$               \\
\end{tabular}
\end{center}
With respect to the two-phase Navier-Stokes equation,
the error in pressure $\psi$ and pressure gradient $\nabla \psi$
are especially inadvertent:
from a physical point of view, a circular droplet in a velocity field
$\vec{u} = 0$
represents a steady state with minimal energy.
Since the velocity is linked to the pressure gradient by
the momentum equation (\ref{eq:Momentum}),
a non-zero pressure gradient would induce an artificial, non-zero
velocity into a state-state solution.

On the other hand, the quadratic  test-case basically confirms that the
extended DG discretization of the Poisson problem is solved
exactly by the numerics, up to machine accuracy.
A necessary perquisite for this is, that that the quadrature rules
used to implement the extended DG discretization are sufficiently precise,
so that the discretization error is not dominated by
the error of the numerical integration.

So, this example demonstrates that the result of
a level-set based two-phase computation may heavily depend on the
choice of the level-set field. The purpose of this study
is to identify filters by which this undesirable dependence can be minimized.

\paragraph{Remark on smooth interface methods:}
those methods usually treat surface tension
as a smoothed Delta-distribution, therefore they do not require $\varphi \in \Cnull(\Omega)$.
However, the discussion on smoothed interface methods is beyond the scope of this publication.

\section{Patch recovery in the $L^2$ -- sense.}
\label{sec:patch_recovery}

We introduce the patch-recovery operator
\begin{equation}
   \mathrm{prc}_w^l : \DGP{q}(\GridCells) \rightarrow \DGP{q}(\GridCells)
\end{equation}
which is used to filter the level-set field $\varphi$ and its derivatives.
Some auxiliary definitions are required:
\begin{itemize}
\item
A cell $K \in \GridCells$ is called to be cut (by the zero-level-set $\frakI$)
if
$ \oint_{K \cap \frakI} 1 \dS \neq 0 $.

\item
A cell $L \in \GridCells$ is called to be a neighbor of cell $K \in \GridCells$,
if they share at least one point, i.e. if
$\overline{L} \cap \overline{K} \neq \{ \}$.
(Rem.: by defining neighbors as sharing one point -- instead of the usual definition of sharing an edge --
this e.g. induces that in a Cartesian grid a quadrilateral element has 8 neighbors, and not just 4
as it would be the case with the usual definition of neighbors.)

\item
 We further define the set of all cut-cells,
\[
  \GridCells^{\mathrm{cc},0} := \{ K \in \GridCells; \ K \textrm{ is cut} \}
\]
and all cut cells and their neighbors
\[
  \GridCells^{\mathrm{cc},1} := \{ K \in \GridCells; \ K \textrm{ or one of its neighbors are cut} \}
\]
and the union of these cells, i.e.
\[
  \Omega^{\mathrm{cc},0} := \Omega^{\mathrm{cc}} := \bigcup_{K \in \GridCells^{\mathrm{cc},0}} K
  \textrm{ and }
  \Omega^{\mathrm{cc},1} := \bigcup_{K \in \GridCells^{\mathrm{cc},1}} K .
\]
\end{itemize}
The patch-recovery operator with width $w \in \{0, 1\}$ is defined as
the $L^2$-projection onto the
broken polynomial space on the composite cell $Q_K$,
 which is formed from all neighbor cells of some cell $K$.
Precisely:
For a cell $K \in \GridCells^{\textrm{cc}, w}$
let be
$L_1, \ldots, L_I \in \GridCells^{\textrm{cc}, w} $
the neighbors of $K$ in $\GridCells^{\textrm{cc}, w}$.
Then one can define a composite cell
$Q_K := K \cup L_1 \cup \ldots \cup L_I$
and the polynomial space of order $q$ on this composite cell,
$\DGP{q}( \{ Q_K \} )$.
For cell $K$, the patch-recovery operation $ \mathrm{prc}_w (u) =: v$ is then defined
as the $L^2$--projection of $u$ onto $\DGP{q}( \{ Q_K \} )$, i.e.
\[
    v|_K := \left\{
    \begin{array}{ll}
    \left. \proj{\DGP{q}( \{ Q_K \} )}{u|_{Q_K} } \right|_K
       & \textrm{if } K \in    \GridCells^{\textrm{cc}, w} \\
    0  & \textrm{if } K \notin \GridCells^{\textrm{cc}, w} \\
    \end{array}
    \right.
    .
\]
By $\mathrm{prc}_w^l (u)$ we denote the composition of $l$ patch-recovery operations,
and consequently $\mathrm{prc}_w^0 (u) = u$.

\paragraph{Implementation of the patch-recovery operator:}
For the space $\DGP{q}(\GridCells)$, we assume
an orthonormal basis $(\phi_{j,n})_{j=1,\ldots,J,n=1,\ldots,N}$,
with $\mathrm{supp}(\phi_{jn}) = \overline{K_j}$.
Here $j$ is called the cell-index, while $n$ is called the mode index.
Then, one chooses a polynomial basis $(\theta_n)_{n=1,\ldots,N}$,
on the composite cell $Q_K$ and computes factors
$A_{nim}$ to express the basis functions $\theta_n$ in terms of $\phi_{jn}$,
i.e.
\[
  \theta_n = \sum_{i,m} \phi_{{l_i} m} A_{n i m}.
\]
$l_i$ denotes a mapping from indices $i$ to the cell indices of the cells which make up $Q_K$.
To aid numerical stability, the basis $\theta_n$ may be ``pre-orthonormalized'', e.g.
the polynomials may be chosen to be orthonormal in the bounding box of $Q_K$.
For the mass matrix $M$ of $\theta_n$, one gets
\begin{alignat}{2}
M_{nm}
& := \int_{Q_K} \theta_n \theta_m \dV \nonumber \\
& = \sum_{i,s} \sum_{j,r} A_{nis} A_{mjr} \underbrace { \int_{\Omega} \phi_{{l_i}s} \phi_{{l_j} r} \dV
    }_{= \delta_{ij} \delta{sr} }  \nonumber \\
& = \sum_{i,r} A_{nir} A_{mir}. \nonumber
\end{alignat}
Then, an orthonormal basis on $Q_K$ is given by
\begin{equation*}
\vartheta_m
= \sum_{n} \theta_m S_{nm}
= \sum_{i,r} \underbrace{ \left(
   \sum_{n} S_{nm} A_{nir}
 \right) }_{ =: B_{mir} } \phi_{{l_i} r}
\end{equation*}
where the change-of-basis matrix $S$ is given by the relation
\begin{equation*}
  S^T M S = I,
  \textrm{ resp. }
  M = (S^{-1})^{T} S^{-1},
\end{equation*}
i.e. $S$ is the inverse of the Cholesky-factor of $M$.
Then, for some function $u = \sum_{j,s} \hat{u}_{{l_j} s} \phi_{{l_j} s}$, the $L^2$-projection
of $u$ onto $\DGP{q}( \{ Q_K \} )$ is given as $v := \sum_{n} \hat{v}_n \vartheta_n$,
with
\begin{alignat}{2}
\hat{v}_m
& = \int_{Q_K} u \vartheta_m  \dV
\nonumber \\
& = \sum_{j,s} \sum_{i,r} \hat{u}_{{l_j} s} B_{m i r} \underbrace{
     \int_{Q_K} \phi_{{l_i} r}  \phi_{{l_j} s} \dV
}_{ = \delta_{i j} \delta_{r s} }
\nonumber \\
& = \sum_{j,s} \hat{u}_{l_j} B_{m j s}. \nonumber
\end{alignat}
Finally, $v$ can be re-expressed in the `original' basis $\phi_{j n}$ of
the space $\DGP{q}(\GridCells)$, within e.g. cell $K_{l_1}$ as
\begin{equation*}
v|_{K_{l_1}} =
\sum_{r} \left( \sum_{m} \hat{v}_m B_{m 1 r} \right) \phi_{{l_1} r}.
\end{equation*}
The computationally most expensive operation is
the construction of the tensors $A_{nir}$, $S_{nm}$ and finally $B_{mir}$.
Once these are computed, multiple evaluations of the patch-recovery
operator are comparatively cheap, since they can be implemented as matrix-vector multiplications.

\section{The configurable curvature algorithm and Results}

\subsection{Test setup}
\label{sec:test_setup}

\begin{figure}
\begin{center}
\begin{tabular}{ccc}
case (a) & case (b) & case (c) \\
`large circle' & `small circle' & `peanut' \\
 &  & \\
\centering
\raisebox{-0.5\height}{
\begin{overpic}[width=0.3\textwidth
]{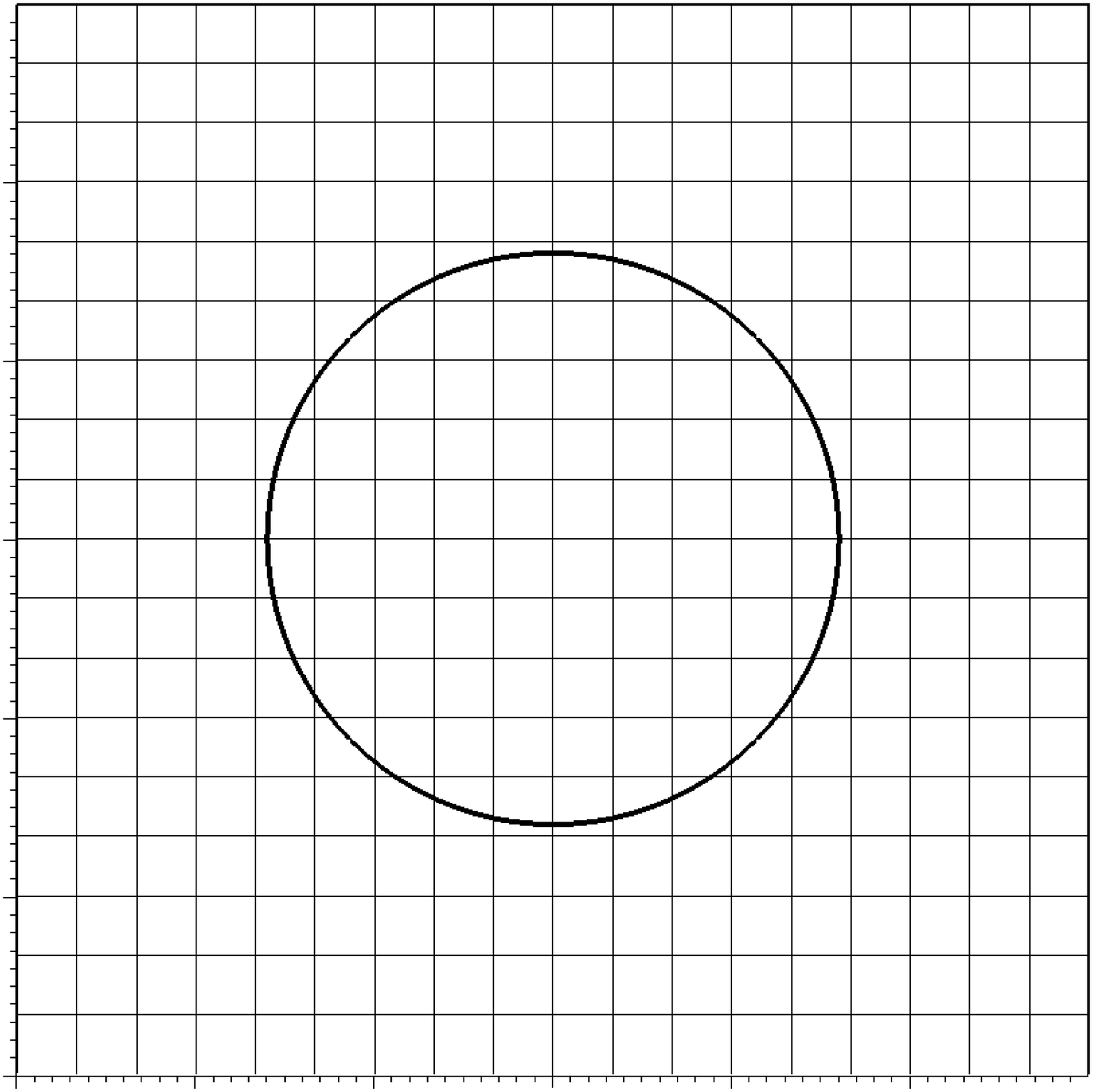}
\put(-5,-7)   {\footnotesize{$-2$}}
\put(48.5,-7) {\footnotesize{$0$}}
\put(98,-7)   {\footnotesize{$2$}}
\put(-12,-2)  {\footnotesize{$-2$}}
\put(-6,48)   {\footnotesize{$0$}}
\put(-6,98)   {\footnotesize{$2$}}
\end{overpic}}
\hspace*{2mm} 
&
\raisebox{-0.5\height}{
\begin{overpic}[width=0.3\textwidth
]{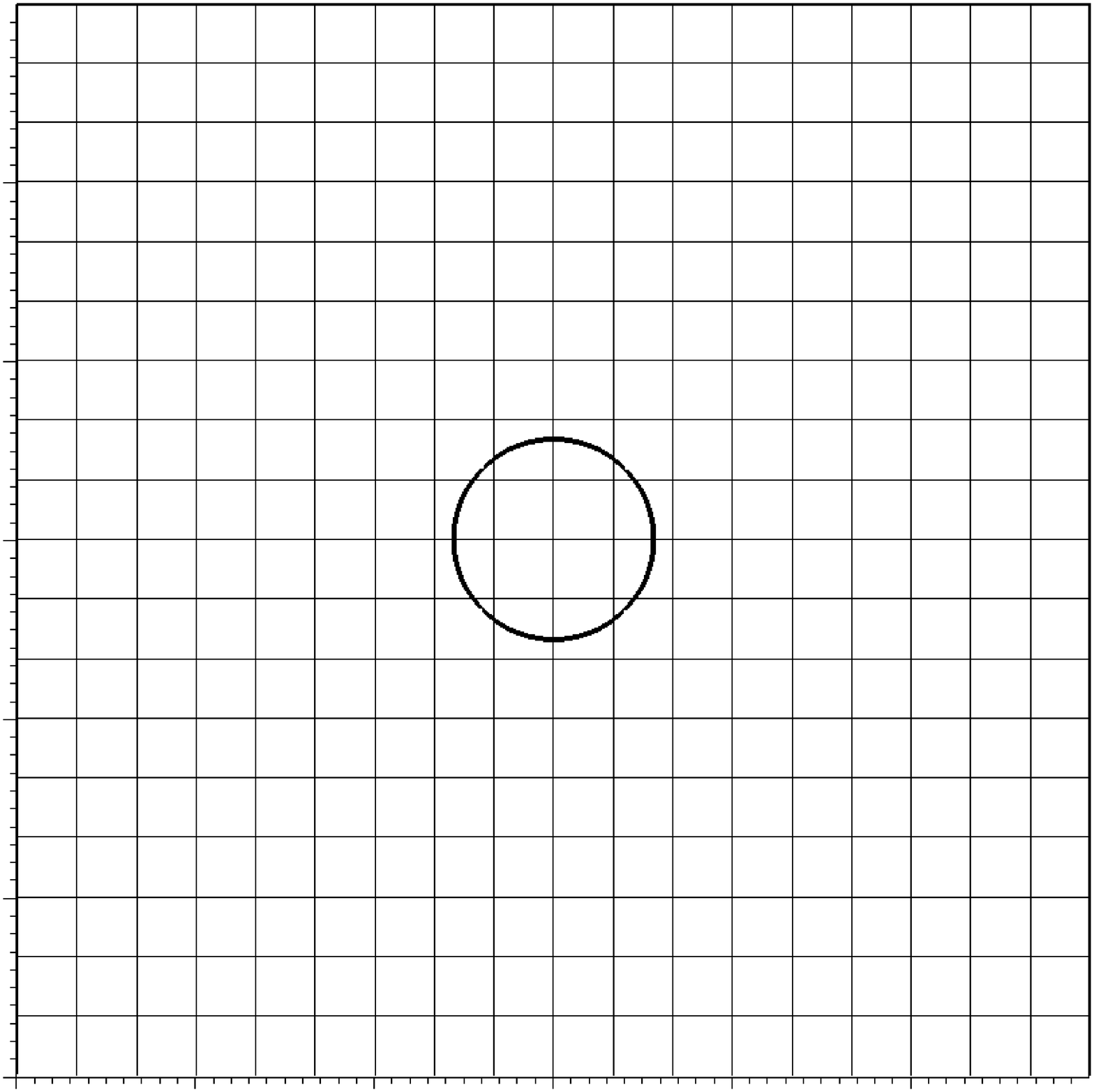}
\put(-5,-7)   {\footnotesize{$-2$}}
\put(48.5,-7) {\footnotesize{$0$}}
\put(98,-7)   {\footnotesize{$2$}}
\put(-12,-2)  {\footnotesize{$-2$}}
\put(-6,48)   {\footnotesize{$0$}}
\put(-6,98)   {\footnotesize{$2$}}
\end{overpic} }
\hspace*{2mm} 
&
\raisebox{-0.5\height}{
\begin{overpic}[width=0.3\textwidth
]{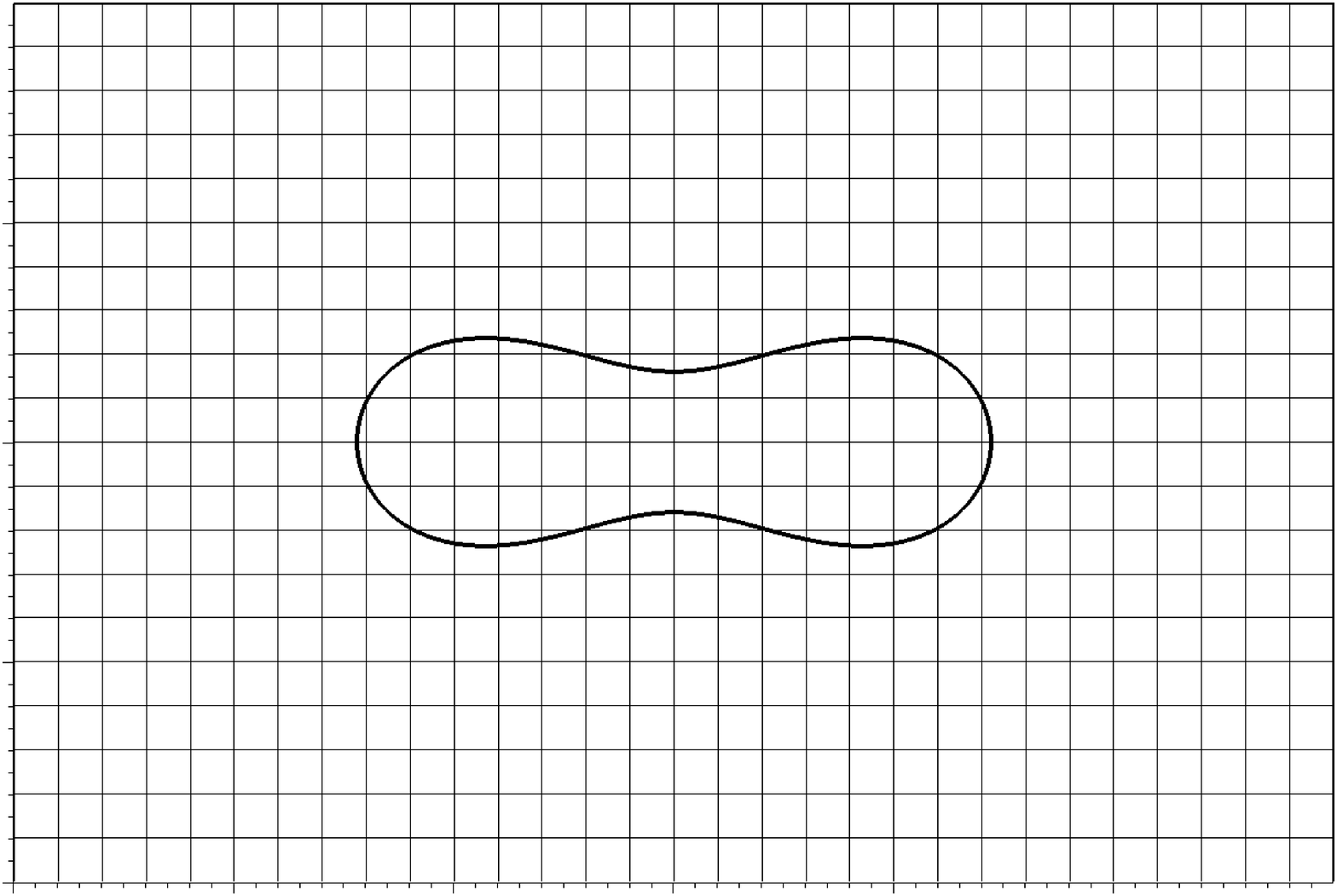}
\put(-5,-7)   {\footnotesize{$-3$}}
\put(48.5,-7) {\footnotesize{$0$}}
\put(98,-7)   {\footnotesize{$3$}}
\put(-12,-2)  {\footnotesize{$-2$}}
\put(-6,31)   {\footnotesize{$0$}}
\put(-6,64)   {\footnotesize{$2$}}
\end{overpic} }
\\
\end{tabular}
\end{center}
  \caption{The three different test geometries which are investigated:
   a rather large circle, with respect to grid width $h$,
   a small circle and a 'peanut'-shaped interface.}
  \label{fig:test-geom}
\end{figure}

We investigate three basic cases, also shown in figure \ref{fig:test-geom};
these are:
\begin{equation*}
\begin{array}{ll}
\textrm{case a, large circle:}    &
     \varphi_{\textrm{ex}} := 0.8 - \sqrt{x^2 + y^2}, \\
\textrm{case b, small circle:}    &
     \varphi_{\textrm{ex}} := 0.25 - \sqrt{x^2 + y^2}, \\
\textrm{case c, 'peanut'-shape:}  &
     \varphi_{\textrm{ex}} :=
     3 - 0.9 \cos(x) - \sqrt{(x+1)^2 + y^2} \\
     & \qquad \qquad \qquad \qquad \qquad - \sqrt{ (x-1)^2 + y^2 }. \\
\end{array}
\end{equation*}
For case (a) and (b), the domain is chosen as $\Omega = (-3/2,3/2)$,
discretized by $18 \times 18$ equidistant cells,
while for case (c) $\Omega = (-3,3)\times(-2,2)$, discretized by $30 \times 20$ equidistant cells.
Since we investigate a huge variety of configurations for the curvature algorithm
-- namely 6144 configurations for each test-case, see below -- we do not
alter the DG polynomial degree of the level-set function nor
of pressure $\psi$ and the pressure gradient.
However, the influence of polynomial degree of the curvature approximation $\kappa$
is investigated, see below.
We define
the approximation of the exact level-set field $\varphi_{\textrm{ex}}$
on the broken and the continuous piecewise polynomial space, i.e.
\begin{equation}
    \varphi_{\textrm{br}} := \proj{\DGP{4}(\GridCells)} { \varphi_{\textrm{ex}} }
\end{equation}
and
\begin{equation}
    \varphi_{\mathcal{C}^0} := \proj{\SEM{2}(\GridCells)} {\varphi_{\textrm{br}}}.
\end{equation}
For each test case and algorithm configuration,
three different $L^2$ errors are recorded:
\begin{equation*}
\begin{array}{ll}
\textrm{curvature on cut-cells:} & \| \kappa - \kappa_{\textrm{ex}} \|_{L^2(\Omega^{\textrm{cc}})} \\
\textrm{pressure:}               & \| \psi - \psi_{\textrm{ex}} \|_{L^2(\Omega)}  \\
\textrm{pressure gradient jump:} & \| \jump{\nabla \psi - \nabla \psi_{\textrm{ex}} } \cdot \nI \|_{L^2(\frakI)}
                                 = \| \jump{\nabla \psi  }  \cdot \nI \|_{L^2(\frakI)} \\
\end{array}
\end{equation*}
The latter error is especially important in the context of a
two-phase Navier-Stokes - problem, as already noted in the first example (section \ref{sec:motivation_example}).
For a circular interface
-- which represents, form a physical point of view, a natural minimal-energy state of a droplet --
one usually wants the pressure gradient $\nabla \psi$ as low as possible.
A non-zero pressure gradient would correspond to some `artificial' velocity
that is introduced in the momentum equation (\ref{eq:Momentum}), preventing the simulation from
reaching the natural, minimal-energy final state with zero velocity.

\paragraph{Evaluation of curvature:}
When applying patch-recovery construction to curvature-evaluation,
several options are at hand:
Where in the algorithm should patch-recovery be applied?
Onto the input, i.e. the level-set function, onto the output,
i.e. the curvature, onto intermediate results like gradients or Hessians,
or everywhere?
Which polynomial degree should be chosen for the curvature?
How many cells around the interface should
be used for
the  patch recovery operation?

\begin{figure}
    \newcommand{\fsrc}{{\small{\LevelSetSource }}}
    \newcommand{\fCN}{\fromCnull}
    \newcommand{\fDG}{\fromDG}

    \newcommand{\f}{$f$}
    \newcommand{\tf}{$\tilde{f}$}
    \newcommand{\g}{$\vec{g}$}
    \newcommand{\tg}{$\tilde{\vec{g}}$}
    \newcommand{\He}{$\vec{H}$}
    \newcommand{\tH}{$\tilde{\vec{H}}$}

    \newcommand{\gsrc}{{\small{\GradientOption}}}
    \newcommand{\Nf}{{\footnotesize{\LevSet}}}
    \newcommand{\Ntf}{{\footnotesize{\FiltLevSet}}}

    \newcommand{\Hsrc}{{\small{\HessianOption}}}
    \newcommand{\Pf}{{\footnotesize{\LevSetH}}}
    \newcommand{\Ng}{{\footnotesize{\LevSetGrad}}}
    \newcommand{\Ptf}{{\footnotesize{\FiltLevSetH}}}
    \newcommand{\Ntg}{{\footnotesize{\FiltLevSetGrad}}}

    \newcommand{\Pra}{$\mathrm{prc}^{l_1}_w$}
    \newcommand{\filG}{{\small{\useFiltLevSetGrad}}}
    \newcommand{\filH}{{\small{\useFiltLevSetHess}}}
    \newcommand{\tka}{$\tilde{\kappa}$}
    \newcommand{\ka}{$\kappa$}
    \newcommand{\Prb}{$\mathrm{prc}^{l_2}_w$}
  \begin{center}
  \input{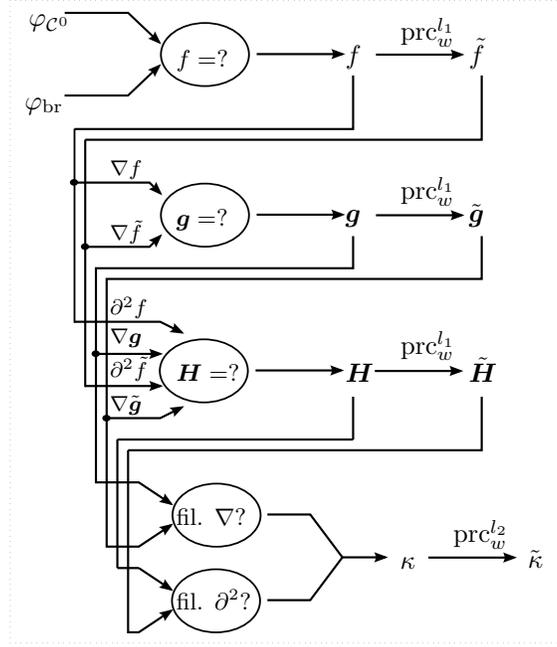} \\
  \end{center}
  \caption{
  Flow chart for curvature evaluation: the filtered curvature $\tilde{\kappa}$
  is computed either from the SEM (\fromCnull) or the DG (\fromDG) representation of the level-set field.
  The configuration of the algorithm depends on the switches for the
  level-set source (\LevelSetSource), for the gradient source (\GradientOption), for the Hessian source (\HessianOption),
  on the number of patch-recovery cycles for the derivatives ($l_1$) and the curvature itself ($l_2$),
  as well as the width $w$ of the patch-recovery domain;
  furthermore, there are switches (\useFiltLevSetGrad, \useFiltLevSetHess) to evaluate Bonnet's formula from
  either the un-filtered gradient $\vec{g}$ or the filtered gradient $\tilde{\vec{g}}$
  and, independently from
  either the un-filtered Hessian $\vec{H}$ or the filtered Hessian $\tilde{\vec{H}}$.
  In addition, not shown in this graph, we vary the DG polynomial
  degree of $\tilde{f}$, $\tilde{\vec{g}}$, $\tilde{\vec{H}}$, $\kappa$ and $\tilde{\kappa}$
  as well as the domain of the patch-recovery.
  }
  \label{fig:flowchart}
\end{figure}

In order to evaluate different configurations for patch recovery -- based curvature algorithms,
a modular algorithm with the following options was implemented; the filtered curvature $\tilde{\kappa}$,
used as an input to the Poisson problem (\ref{eq:Poisson}), resp. (\ref{eq:XDGPoissonSys}),
is computed as
\begin{equation}
\left[ \begin{array}{ll}
f := &
  \textrm{one of} \left\{ \begin{array}{l} \varphi_{\mathcal{C}^0} \\ \varphi_{\mathrm{br}} \end{array} \right. \\
\tilde{f} := &
  \mathrm{prc}_w^{l_1} (f) \\
\vec{g} := &
  \textrm{one of} \left\{ \begin{array}{l} \nabla f \\ \nabla \tilde{f} \end{array} \right. \\
\tilde{\vec{g}} := &
  \mathrm{prc}_w^{l_1} (\vec{g}) \\
\vec{H} := &
  \textrm{one of} \left\{ \begin{array}{l}
          \partial^2 f \\
          \nabla \vec{g} \\
          \partial^2 \tilde{f} \\
          \nabla \tilde{\vec{g}} \\
  \end{array} \right. \\
\tilde{\vec{H}} := &
  \mathrm{prc}_w^{l_1} (\vec{H}) \\
\kappa := &
  \mathrm{curv} \left(
  \textrm{one of} \left\{ \begin{array}{l} \vec{g} \\ \tilde{\vec{g}} \end{array} \right. ,
  \textrm{one of} \left\{ \begin{array}{l} \vec{H} \\ \tilde{\vec{H}} \end{array} \right.
     \right) \\
\tilde{\kappa} := &  \mathrm{prc}_w^{l_2} (\kappa) \\
\end{array} \right.
\end{equation}
A flowchart is given in figure \ref{fig:flowchart}. In detail, this algorithm provides
the following configuration options:
\begin{itemize}
  \item Which field should be used as an input for the algorithm (switch `\LevelSetSource')?
        \begin{itemize}
           \item the SEM-representation (case `\fromCnull'), i.e. $f := \varphi_{\mathcal{C}^0}$.
           \item the DG-representation (case `\fromDG'), i.e. $f := \varphi_{\mathrm{br}}$.
        \end{itemize}
  \item From which field should the level-set gradient $\vec{g}$ be computed (switch `\GradientOption')?
        \begin{itemize}
          \item from the un-filtered level-set (case `\LevSet'), i.e  $\vec{g} := \nabla f$.
          \item from the filtered level-set (case `\FiltLevSet'), i.e $\vec{g} := \nabla \tilde{f}$.
       \end{itemize}
  \item From which field should the Hessian $\vec{H}$ of the level-set be computed (switch `\HessianOption')?
        \begin{itemize}
          \item from the un-filtered level-set (case `\LevSetH'), i.e. $\vec{H} := \partial^2 f$.
          \item from the un-filtered level-set gradient (case `\LevSetGrad'), i.e. $\vec{H} := \nabla \vec{g}$.
          \item from the filtered level-set (case `\FiltLevSetH'), i.e. $\vec{H} := \partial^2 \tilde{f}$.
          \item from the filtered level-set gradient (case `\FiltLevSetGrad'), i.e. $\vec{H} := \nabla \tilde{\vec{g}}$.
       \end{itemize}
  \item The number $l_1 \in \{ 1, 2, 5, 10 \}$
        of patch-recovery-cycles that are applied to the level-set field and its derivatives,
        i.e. $\tilde{f} := \patchrec{l_1}(f)$, $\tilde{\vec{g}} := \patchrec{l_1}(\vec{g})$ and $\tilde{\vec{H}} := \patchrec{l_1}(\vec{H})$.
  \item The number $l_2 \in \{ 0, 1, 5, 10 \}$
        of patch-recovery-cycles that are applied to the computed curvature $\kappa$
        i.e. $\tilde{\kappa} := \patchrec{l_2}(\kappa)$.
  \item The DG-polynomial degree of the curvature $\kappa$ and the filtered derivatives is altered:
      $\kappa, \tilde{\kappa}, \tilde{f} \in \DGP{\alpha p}(\GridCells)$,
      $\tilde{\vec{g}} \in \DGP{\alpha p}(\GridCells)^2$,
      $\tilde{\vec{H}} \in \DGP{\alpha p}(\GridCells)^{2 \times 2}$,
      where $p = 4$ is the polynomial degree of $\varphi_{\mathrm{DG}}$,  with the integer multiplier  $\alpha \in \{ 1, 2, 3 \}$.
\end{itemize}

\subsection{Results and discussion}
\label{sec:results_and_discussion}

For all test-cases, the three different errors measures span across several magnitudes,
see figures \ref{fig:scatterCurv} and \ref{fig:scatterPoisson}, with one
exception, namely the error
$\| \psi - \psi_{\textrm{ex}} \|_{L^2(\Omega)}$
in case (c), see bottom-left plot in figure \ref{fig:scatterPoisson}.
Therefore, the pressure error of case (c) is not considered in the survey.

\begin{itemize}
\item
It is, in all test-cases generally better
to use the DG-representation ($f = \varphi_{\textrm{br}}$) for
curvature computation than the continuous CG ($f = \varphi_{\mathcal{C}^0}$).
Considering that the broken approximation $\varphi_{\textrm{br}} \in \DGP{4}(\GridCells)$
is in general more precise than the continuous approximation $\varphi_{\mathcal{C}^0} \in \SEM{2}(\GridCells)$,
this seems not very surprising:
\begin{center}
\begin{tabular}{l||c|c}
     & $\| \varphi_{\textrm{br}} -  \varphi_{\textrm{ex}} \|_{L^2(\Omega^{\textrm{cc}})}$
     & $\| \varphi_{\mathcal{C}^0} - \varphi_{\textrm{ex}} \|_{L^2(\Omega^{\textrm{cc}})}$
     \\ \hline
case (a) & $3.192 \cdot 10^{-7}$ & $5.087 \cdot 10^{-5}$  \\
case (b) & $1.165 \cdot 10^{-5}$ & $2.396 \cdot 10^{-4}$  \\
case (c) & $1.71 \cdot 10^{-5} $ & $3.759 \cdot 10^{-4}$  \\
\multicolumn{3}{c}{} \\
     & $\| \nabla \varphi_{\textrm{br}} -  \nabla \varphi_{\textrm{ex}} \|_{(L^2(\Omega^{\textrm{cc}}))^2}$
     & $\| \nabla \varphi_{\mathcal{C}^0} - \nabla \varphi_{\textrm{ex}} \|_{(L^2(\Omega^{\textrm{cc}}))^2}$
     \\ \hline
case (a) & $3.128 \cdot 10^{-5}$ & $2.008 \cdot 10^{-3}$   \\
case (b) & $1.096 \cdot 10^{-3}$ & $1.096 \cdot 10^{-2}$  \\
case (c) & $1.342 \cdot 10^{-3}$ & $1.418 \cdot 10^{-2}$  \\
\multicolumn{3}{c}{}     \\
     & $\| \partial^2 \varphi_{\textrm{br}} -  \partial^2 \varphi_{\textrm{ex}} \|_{(L^2(\Omega^{\textrm{cc}}))^{2 \times 2}}$
     & $\| \partial^2 \varphi_{\mathcal{C}^0} - \partial^2 \varphi_{\textrm{ex}} \|_{(L^2(\Omega^{\textrm{cc}}))^{2 \times 2}}$
     \\ \hline
case (a) & $2.173 \cdot 10^{-3}$ & $9.045 \cdot 10^{-2}$  \\
case (b) & $7.806 \cdot 10^{-2}$ & $4.364 \cdot 10^{-1}$   \\
case (c) & $7.958 \cdot 10^{-2}$ & $4.71 \cdot 10^{-1} $ \\
\end{tabular}
\end{center}

The minimal errors achieved for the broken and the continuous approximation, over all configurations are:
\begin{center}
\begin{tabular}{l||c|c}
     & min. over configs. with       & min. over configs. with         \\
     & broken approx.                & continuous approx.              \\
     & ($f = \varphi_{\textrm{br}}$) & ($f = \varphi_{\mathcal{C}^0}$) \\
\hline
\multicolumn{3}{c}{\rule[-7pt]{0mm}{19pt} curvature error $\| \kappa - \kappa_{\textrm{ex}} \|_{L^2(\Omega^{\textrm{cc}})}$: } \\
\hline
case (a) & $1.874 \cdot 10^{-5}$  & $8.008 \cdot 10^{-4}$ \\
case (b) & $3.821 \cdot 10^{-3}$  & $3.606 \cdot 10^{-2}$ \\
case (c) & $5.46  \cdot 10^{-3}$  & $6.43  \cdot 10^{-2}$ \\
\hline
\multicolumn{3}{c}{\rule[-7pt]{0mm}{19pt} pressure error $\| \psi - \psi_{\textrm{ex}} \|_{L^2(\Omega)}$}: \\
\hline
case (a) & $1.053 \cdot 10^{-6}$  & $1.697 \cdot 10^{-5}$ \\
case (b) & $2.669 \cdot 10^{-5}$  & $6.34  \cdot 10^{-5}$ \\
case (c) & $1.403 \cdot 10^{-1}$  & $1.542 \cdot 10^{-1}$ \\
\hline
\multicolumn{3}{c}{\rule[-7pt]{0mm}{19pt} pressure gradient jump error $ \| \jump{\nabla \psi  }  \cdot \nI \|_{L^2(\frakI)}$}: \\
\hline
case (a) & $1.928 \cdot 10^{-4}$  & $4.19  \cdot 10^{-4}$ \\
case (b) & $2.519 \cdot 10^{-3}$  & $3.916 \cdot 10^{-3}$ \\
case (c) & $5.966 \cdot 10^{-2}$  & $5.454 \cdot 10^{-2}$ \\
\end{tabular}
\end{center}
Most interestingly, the latter error measure,
$ \| \jump{\nabla \psi  }  \cdot \nI \|_{L^2(\frakI)}$,
seems not to be affected by the choice of broken versus continuous approximation.

\item Regarding performance, the most influential factors are
      -- not surprisingly --
      the polynomial order of
      the curvature approximation space $\DGP{4 \alpha}(\GridCells)$,
      and the width $w$ of the patch-recovery domain.
      The major computational cost of the patch-recovery operator is
      the construction of the projector onto the aggregate cells,
      while the evaluation of the patch-recovery is in comparison quite fast.
      Therefore, the number of patch-recovery cycles $l_1$ and $l_2$
      is only of minor influence to the run-time.
      For all algorithm configurations which employ any kind of patch-recovery,
      we observe the following mean run-times $\mu_{t_{\textrm{run},\alpha,w}}$
      and standard deviations $\sigma_{t_{\textrm{run},\alpha,w}}  $,
      for different $(\alpha,w)$ - pairs:

\begin{center}
\begin{tabular}{ll||c|c|c|c|c|c|l}
         & $(\alpha,w) \rightarrow$      & 1,0  & 1,1  & 2,0 & 2,1 & 3,0 & 3,1  & \\
\hline \hline
case (a) & $\mu_{t_{\textrm{run},\alpha,w}}$
                                         & 0.29 & 0.7  & 1.3 & 6.2 & 9.0 & 48   & seconds \\
         & $\sigma_{t_{\textrm{run},\alpha,w}}  $
                                         & 0.53 & 0.84 & 1.1 & 2.5 & 3.0 & 6.9  & seconds \\
         & $\left( \frac{\mu_{t_{\textrm{run},\alpha,w}}}{\mu_{t_{\textrm{run},1,0}}} \right) $
                                         & 1.0  & 2.5  & 4.5 & 22  & 32  & 168  & normalized \\
         & $\left( \frac{\sigma_{t_{\textrm{run},\alpha,w}}}{\mu_{t_{\textrm{run},1,0}}} \right)$
                                         & 1.9  & 2.9  & 4.0 & 8.7 & 10  & 24   & normalized \\
  \hline
case (b) & $\mu_{t_{\textrm{run},\alpha,w}}$
                                         & 0.24 &   0.38 &  0.62 & 2.4 & 3.9 & 17  & seconds \\
         & $\sigma_{t_{\textrm{run},\alpha,w}}$
                                         & 0.49 &   0.62 &  0.79 & 1.5 & 2.0 & 4.1 & seconds \\
         & $\left( \frac{\mu_{t_{\textrm{run},\alpha,w}}}{\mu_{t_{\textrm{run},1,0}}} \right)$
                                         & 1.0  &   1.6  &  2.6  & 9.7 & 16  & 71  & normalized \\
         & $\left( \frac{\sigma_{t_{\textrm{run},\alpha,w}}}{\mu_{t_{\textrm{run},1,0}}} \right)$
                                         & 2.0  &   2.5  &  3.2  & 6.3 & 8.1 & 17  & normalized \\
  \hline
case (c) & $\mu_{t_{\textrm{run},\alpha,w}}$
                                         & 0.25 &  0.62 & 1.3 & 7.6  & 11   & 63  & seconds \\
         & $\sigma_{t_{\textrm{run},\alpha,w}}  $
                                         & 0.50 &  0.79 & 1.2 & 2.7  & 3.4  & 8.0 & seconds \\
         & $\left( \frac{\mu_{t_{\textrm{run},\alpha,w}}}{\mu_{t_{\textrm{run},1,0}}} \right)$
                                         & 1.0  &  2.5  & 5.4 &  31  &  46  & 257 & normalized \\
         & $\left( \frac{\sigma_{t_{\textrm{run},\alpha,w}}}{\mu_{t_{\textrm{run},1,0}}} \right)$
                                         & 2.0  &  3.2  & 4.7 &  11  &  14  &  32 & normalized \\
\end{tabular}
\end{center}

It becomes apparent that especially the cases with $(\alpha,w) = (3,0)$ and $(\alpha,w) = (3,1)$
are computationally expensive: among all test-cases, the $(\alpha,w) = (3,0)$ -- configurations
are between 16 and 38 times more expensive than the $(\alpha,w) = (1,0)$ -- configuration;
for the $(\alpha,w) = (3,1)$ -- configurations that factor is 71 and 257.

\item
For all test-cases, no significant improvement
int the errors
$\| \psi - \psi_{\textrm{ex}} \|_{L^2(\Omega)}$
and
$\| (\nabla \psi - \nabla \psi_{\textrm{ex}} )\cdot \nI \|_{L^2(\frakI)}$
could be observed
for $\alpha > 2$ and $w > 0$.
Furthermore, no significant improvement could be obtained by
any kind of filtering of Level-Set gradient or Hessian, nor by filtering the curvature itself.
Indeed, from figures \ref{fig:scatterCurv} and \ref{fig:scatterPoisson},
it becomes obvious that
configurations (marked as $\bullet$ in cited figures) which
\begin{itemize}
  \item use the broken polynomial-level-set field ($f = \varphi_{\textrm{br}}$),
  \item perform patch recover only on cut cells itself ($w=0$),
  \item use polynomial degree $2 \cdot p$ for $\tilde{f}$ and $\kappa$  ($\alpha = 2)$,
  \item compute gradient and Hessian from the filtered level-set field
        ($\vec{g} = \nabla \tilde{f}$, $\vec{H} = \partial^2 \tilde{f}$),
  \item employ no additional filtering of the gradient nor the Hessian
        ($\textrm{fil. } \nabla = \mathrm{\false}$, $\textrm{fil. } \partial^2 = \mathrm{\false}$) and
  \item employ no additional filtering of computed curvature ($l_2 = 0$)
\end{itemize}
are among the best-performing and fastest configurations for test cases, with respect to each error measure.

\item
As obvious from figures
\ref{fig:scatterCurv} and \ref{fig:scatterPoisson},
there are certain
configurations which perform better in certain test cases.
However, there are no configurations which perform
significantly better overall, i.e. with
respect to all three error measures in all three test cases.
In order to check that,
we selected the 10 fastest configurations out of the 100
best-performing, with respect to the three different
error measures in all test-cases.
From those configurations, which performed well overall,
none significantly out-performed the configuration recommendation
given above.

\begin{figure}
\begin{center}
\input{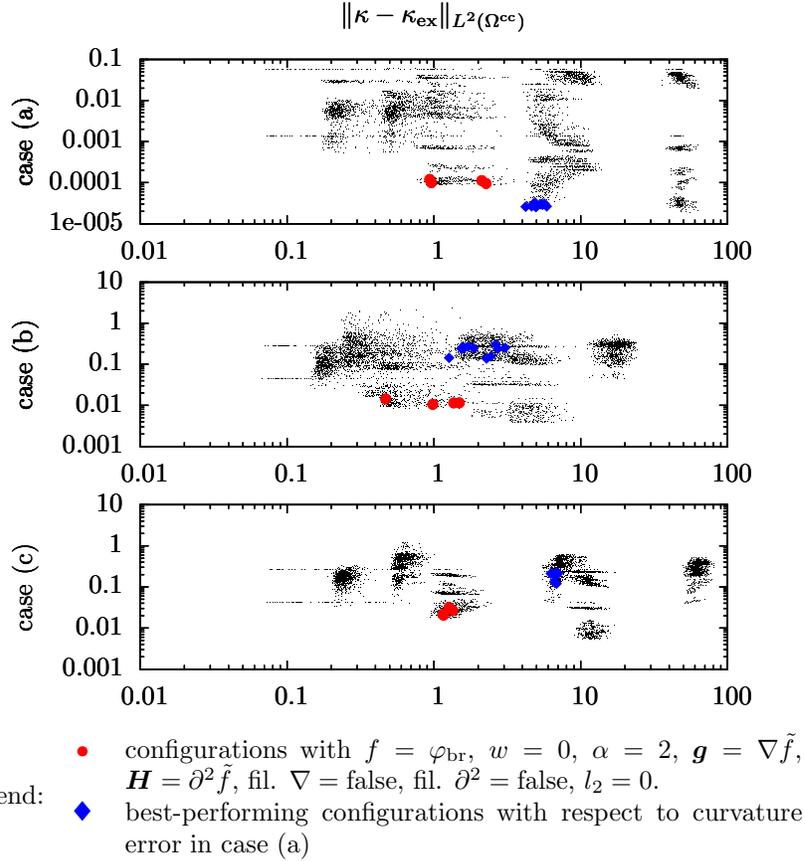} \\
\rule{0mm}{10pt} \\
Legend:
\begin{tabular}{cp{9cm}}
\textcolor{red}{$\bullet$}
   & configurations with
                $f = \varphi_{\textrm{br}}$,
                $w=0$, $\alpha = 2$,
                $\vec{g} = \nabla \tilde{f}$,
                $\vec{H} = \partial^2 \tilde{f}$,
                $\textrm{fil. } \nabla = \mathrm{\false}$, $\textrm{fil. } \partial^2 = \mathrm{\false}$,
                $l_2 = 0$. \\
\textcolor{blue}{$\vardiamond$}
   & best-performing configurations with respect to curvature error in case (a) \\
\end{tabular}
\end{center}
  \caption{
A scatter plot, showing all algorithm configurations, for all test cases,
presenting the runt-time of the curvature evaluation
versus the error measure
$\| \kappa - \kappa_{\textrm{ex}} \|_{L^2(\Omega^{\textrm{cc}})}$.
Configurations marked with symbol $\bullet$
are overall well-performing, i.e. they are among the most accurate and fastest
configurations for each of the three error measures and all test cases, see also figure \ref{fig:scatterPoisson}.
There are certainly configurations, like the ones marked by $\vardiamond $ ,
which perform better in a specific test-case, but not overall.
}
\label{fig:scatterCurv}
\end{figure}

\begin{figure}
\begin{center}
\input{scatter_PoissonErr_C0.tex}
\rule{0mm}{10pt} \\
Legend:
\begin{tabular}{cp{9cm}}
\textcolor{red}{$\bullet$}
   & configurations with
                $f = \varphi_{\textrm{br}}$,
                $w=0$, $\alpha = 2$,
                $\vec{g} = \nabla \tilde{f}$,
                $\vec{H} = \partial^2 \tilde{f}$,
                $\textrm{fil. } \nabla = \mathrm{\false}$, $\textrm{fil. } \partial^2 = \mathrm{\false}$,
                $l_2 = 0$. \\
\end{tabular}
\end{center}
\caption{
A scatter plot, showing all algorithm configurations, for all test cases,
presenting the run-time of the curvature evaluation
versus the two different error measures
$\| \psi - \psi_{\textrm{ex}} \|_{L^2(\frakI)}$
and
$\| \jump{\nabla p} \|_{L^2(\frakI)}$.
Configurations marked with symbol $\bullet$
are overall well-performing, i.e. they are among the most accurate and fastest
configurations for each of the three error measures and all test cases, see also figure \ref{fig:scatterCurv}.
}
  \label{fig:scatterPoisson}
\end{figure}

\item
Finally, we perform a separate discussion of those configurations which use the
continuous representation of the level-set field ($f = \varphi_{\mathcal{C}^0}$).
While $\varphi_{\mathcal{C}^0}$ itself is continuous,
its first and second derivatives contain higher levels of oscillations
than the broken polynomial representation $\varphi_{\textrm{br}}$ of the level-set field,
as already noted above.
Here, filtering of the gradient $\vec{g}$ and the Hessian $\vec{H}$
provide a significant improvement.
The top-5 configurations overall are:
\begin{center}
\begin{tabular}{c|c|c|c|c|c|c|c}
\GradientOption & \HessianOption & \useFiltLevSetGrad & \useFiltLevSetHess & $\alpha$ & $w$ & $l_1$ & $l_2$  \\
\hline
\FiltLevSet & \FiltLevSetGrad & \true  & \true  & 1 & 0 & 1 & 0 \\
\FiltLevSet & \FiltLevSetGrad & \true  & \true  & 1 & 0 & 1 & 1 \\
\FiltLevSet & \FiltLevSetGrad & \true  & \false & 1 & 0 & 1 & 1 \\
\FiltLevSet & \FiltLevSetGrad & \false & \false & 1 & 0 & 1 & 1 \\
\FiltLevSet & \FiltLevSetGrad & \false & \false & 1 & 0 & 2 & 1 \\
\end{tabular}
\end{center}
Common for all of these cases is that
the gradient $\vec{g}$ is compute from the filtered level $\tilde{f}$,
the Hessian  $\vec{H}$ is computed from the filtered gradient $\tilde{\vec{g}}$,
the best results were quite surprisingly achieved for lower order approximation spaces
(i.e. $\alpha = 1$, i.e.
$\tilde{f}, \vec{g}, \tilde{\vec{g}}, \vec{\tilde{H}}, \kappa, \tilde{\kappa}$
are of DG-polynomial order $1 \cdot \alpha$)
and that the width of the patch-recovery domain $w=0$.
The performance of these top-5 configuration is visualized in the scatter plots
figure \ref{fig:scatterCurvCnull} and \ref{fig:scatterPoissonCnull}.

\begin{figure}
\begin{center}
\input{scatter_CurvErr_C0.tex} \\
\rule{0mm}{10pt} \\
Legend:
\begin{tabular}{cp{9cm}}
\textcolor{red}{$\bullet$}
  & top-5 configurations overall; common configuration settings are:
            $\vec{g} = \nabla \tilde{f}$, $\vec{H} = \nabla \tilde{\vec{g}}$, $\alpha = 1$, $w = 0$. \\
\end{tabular}
\end{center}
  \caption{
A scatter plot, showing all algorithm configurations which use the
continuous representation of the level-set ($f = \varphi_{\mathcal{C}^0}$), for all test cases,
presenting the runt-time of the curvature evaluation
versus the error measure
$\| \kappa - \kappa_{\textrm{ex}} \|_{L^2(\Omega^{\textrm{cc}})}$.
Configurations marked with symbol $\bullet$
are overall well-performing, i.e. they are among the most accurate and fastest
configurations for each of the three error measures and all test cases,
see also figure \ref{fig:scatterPoissonCnull}.
There are certainly configurations, like the ones marked by $\times$,
which perform better in a specific test-case, but not overall.
}
\label{fig:scatterCurvCnull}
\end{figure}

\begin{figure}
\begin{center}
\input{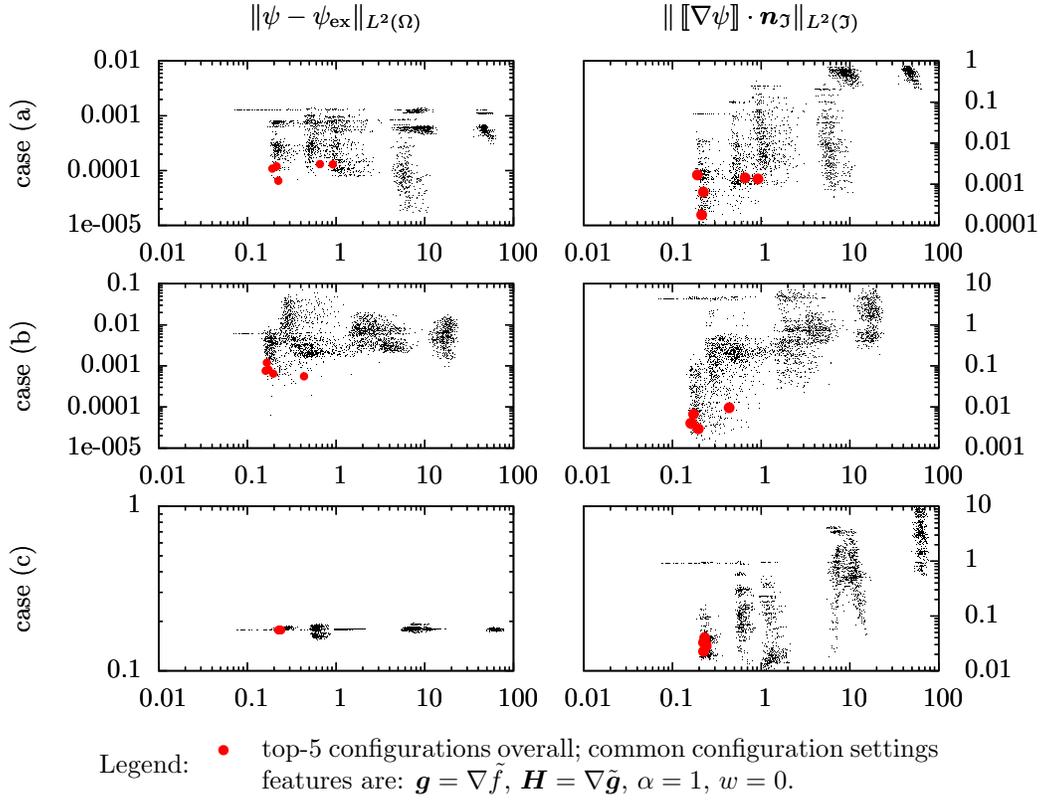}
\rule{0mm}{10pt} \\
Legend:
\begin{tabular}{cp{9cm}}
\textcolor{red}{$\bullet$}
  & top-5 configurations overall; common configuration settings features are:
            $\vec{g} = \nabla \tilde{f}$, $\vec{H} = \nabla \tilde{\vec{g}}$, $\alpha = 1$, $w = 0$. \\
\end{tabular}
\end{center}
\caption{
A scatter plot, showing all algorithm configurations which use the
continuous representation of the level-set ($f = \varphi_{\mathcal{C}^0}$), for all test cases,
presenting the run-time of the curvature evaluation
versus the two different error measures
$\| \psi - \psi_{\textrm{ex}} \|_{L^2(\frakI)}$
and
$\| \jump{\nabla p} \|_{L^2(\frakI)}$.
Configurations marked with symbol $\bullet$
are overall well-performing, i.e. they are among the most accurate and fastest
configurations for each of the three error measures and all test cases,
see also figure \ref{fig:scatterCurvCnull}.
}
  \label{fig:scatterPoissonCnull}
\end{figure}
\end{itemize}

\section{Conclusions and outlook.}
\label{sec:conclusions}

A very important result of this survey is
first,
that we found that there is no need
to increase the polynomial degree of the filtered properties
by more than a factor of two times, in comparison to the polynomial degree
of the original level-set field;
second, that we found it sufficient to perform the
patch recovery just on the layer of cut-cells themselves, and not considering
any values outside of that band.
Since the run-time is dominated by those two factors, it is very pleasant
that these factors can be kept low.

Furthermore, we were able to show the additional benefit of filtering
also the gradient and the Hessian in the case of the spectral-element representation of the
level set.

Within this study we focused on the curvature computation in a very controlled
setup, where an analytic expression for the level-set field was given.
We are currently developing a two-phase Navier-Stokes solver based on the
extended DG discretization presented in section \ref{sec:XDGPoisson}
where curvature computation is an important part.

This survey focused solely on patch recovery for filtering DG-based properties.
Obviously, there other techniques for constructing filters, e.g. WENO,
which we may investigate in future works.

\appendix

\section{Extended DG discretization of the Poisson problem}
\label{sec:XDGPoisson}

Finally, we give a brief description of the extended discontinuous Galerkin method
which is used to solve the Poisson eq. (\ref{eq:Poisson}).
We search for $(\psi,\vec{v}) \in \XDG{2}(\GridCells) \times \XDG{3}(\GridCells)^2$,
so that for all $(f,\vec{w}) \in \XDG{2}(\GridCells) \times \XDG{3}(\GridCells)^2$
\begin{equation}
\begin{array}{rcl}
b(\psi, \vec{w} ) & = & s(\vec{w}) \\
b(f, \vec{w}    ) & = & 0          \\
\label{eq:XDGPoissonSys}
\end{array}
\end{equation}
The bilinear form $b(-,-)$ is given as
\begin{equation*}
  b(f, \vec{w} ) =
      - \int_\Omega f \ \divergence{\vec{w}} \dx
     - \oint_{\Gamma \cup \frakI} \jump{\vec{v}} \cdot \normGammaI \mean {p} \dS.
\end{equation*}
while the linear form $s(-)$, which represents the surface-tension terms,
is given as
\begin{equation*}
  s(\vec{w}) = \oint_{\frakI}
  \sigma \tilde{\kappa} \nI \cdot \jump{ \vec{w} }
  \dS
  .
\end{equation*}
Again, $\nabla$ denotes also the broken gradient, and $\divergence{-}$
the broken divergence.
The set $\Gamma$ is the union of all edges of all cells, i.e.
$\Gamma := \bigcup_{K \in \GridCells} \partial K$
and $\normGammaI$ denotes an almost-everywhere continuous normal field, which is
equal to $\nI$ on $\frakI$, an outer normal on $\partial \Omega$
and normal onto the cell boundaries.
Given that, the jump- and mean-value operator are defined as
\begin{equation*}
\begin{array}{rcll}
\jump{u}(\vec{x}) & := &
 \lim_{\xi \searrow 0} \left(  u(\vec{x} + \xi \normGammaI) - u(\vec{x} - \xi \normGammaI) \right) &
      \textrm{ on } (\Gamma \setminus \partial \Omega ) \cup \frakI, \\
\mean{u}(\vec{x}) & := &
 \lim_{\xi \searrow 0} \frac{1}{2}  \left(  u(\vec{x} + \xi \normGamma) + u(\vec{x} - \xi \normGamma) \right) &
      \textrm{ on } (\Gamma \setminus \partial \Omega ) \cup \frakI, \\
 \jump{u}(\vec{x}) & := & \lim_{{\xi \searrow 0}} u(\vec{x} - \xi \normGammaI) & \textrm{ on } \partial \Omega , \\
 \mean{u}(\vec{x}) & := & \lim_{{\xi \searrow 0}} u(\vec{x} - \xi \normGammaI) & \textrm{ on } \partial \Omega  .\\
\end{array}
\end{equation*}
In order to perform numerical integration on the cut cells, i.e.
on domains like $K \cap \frakA$, $K \cap \frakB$, $K \cap \frakI$, etc.,
a quadrature technique presented in \cite{muller_highly_2013} is used.
The linear system obtained from discretization (\ref{eq:XDGPoissonSys})
is solved by the direct sparse solver PARDISO, see
\cite{schenk_2000,schenk_2004,schenk_2006}.


\bibliography{curvstudy}{}

\begin{thebibliography}{17}
\providecommand{\natexlab}[1]{#1}
\providecommand{\url}[1]{\texttt{#1}}
\expandafter\ifx\csname urlstyle\endcsname\relax
  \providecommand{\doi}[1]{doi: #1}\else
  \providecommand{\doi}{doi: \begingroup \urlstyle{rm}\Url}\fi

\bibitem[Chen et~al.(2004)Chen, Minev, and Nandakumar]{chen_projection_2004}
T.~Chen, P.~D. Minev, and K.~Nandakumar.
\newblock A projection scheme for incompressible multiphase flow using adaptive
  eulerian grid.
\newblock \emph{International Journal for Numerical Methods in Fluids},
  45:\penalty0 1--19, May 2004.
\newblock ISSN 0271-2091, 1097-0363.

\bibitem[Cheng and Fries(2012)]{cheng_xfem_2012}
K.-W. Cheng and T.-P. Fries.
\newblock {XFEM} with hanging nodes for two-phase incompressible flow.
\newblock \emph{Computer Methods in Applied Mechanics and Engineering},
  245-246:\penalty0 290--312, 2012.
\newblock ISSN 00457825.
\newblock \doi{10.1016/j.cma.2012.07.011}.
\newblock URL
  \url{http://linkinghub.elsevier.com/retrieve/pii/S0045782512002319}.

\bibitem[Di~Pietro and Ern(2011)]{di_pietro_mathematical_2011}
D.~A. Di~Pietro and A.~Ern.
\newblock \emph{Mathematical Aspects of Discontinuous Galerkin Methods}.
\newblock Number~69 in {M}ath{\'e}matiques et {A}pplications. Springer, 2011.
\newblock ISBN 9783642229794.

\bibitem[{Gross} and Reusken(2007)]{SvenGross06}
S.~{Gross} and A.~Reusken.
\newblock Finite element discretization error analysis of a surface tension
  force in two-phase incompressible flows.
\newblock \emph{SIAM J. Numer. Anal.}, 45\penalty0 (4):\penalty0 1679--1700,
  2007.

\bibitem[Heimann(2013)]{heimannThesis2013}
F.~Heimann.
\newblock \emph{An Unfitted Higher-Order Discontinuous Galerkin Method for
  Incompressible Two-Phase Flow with Moving Contact Lines}.
\newblock {PhD} thesis, Heidelberg University, Heidelberg, 2013.

\bibitem[Heimann et~al.(2013)Heimann, Engwer, Ippisch, and
  Bastian]{heimann_unfitted_2013}
F.~Heimann, C.~Engwer, O.~Ippisch, and P.~Bastian.
\newblock An unfitted interior penalty discontinuous {G}alerkin method for
  incompressible {Navier-Stokes} two-phase flow.
\newblock \emph{International Journal for Numerical Methods in Fluids},
  71\penalty0 (3):\penalty0 269--293, 2013.
\newblock ISSN 02712091.
\newblock \doi{10.1002/fld.3653}.
\newblock URL \url{http://doi.wiley.com/10.1002/fld.3653}.

\bibitem[Herrmann(2011)]{herrmann_simulating_2011}
M.~Herrmann.
\newblock On simulating primary atomization using the refined level set grid
  method.
\newblock \emph{Atomization and Sprays}, 21\penalty0 (4):\penalty0 283--301,
  2011.
\newblock ISSN 1044-5110.
\newblock \doi{10.1615/AtomizSpr.2011002760}.
\newblock URL
  \url{http://www.begellhouse.com/journals/6a7c7e10642258cc,4fa736af31a9c3d7,32765dc85aa6de1e.html}.

\bibitem[Herrmann and Jibben(2012)]{JibbenHerrmann2012}
M.~Herrmann and Z.~Jibben.
\newblock A {{R}unge-{K}utta} discontinuous {G}alerkin conservative level set
  method.
\newblock In \emph{Center for Turbulence Research, Annual Research Briefs
  2012}, pages 305--314, 2012.
\newblock URL \url{http://ctr.stanford.edu/Summer/SP12/05.01_jibben.pdf}.

\bibitem[Hesthaven and Warburton(2008)]{hesthaven_nodal_2008}
J.~S. Hesthaven and T.~Warburton.
\newblock \emph{Nodal Discontinuous {G}alerkin Methods: Algorithms, Analysis,
  and Applications}.
\newblock Number~54 in Texts in Applied Mathematics. Springer-Verlag, 2008.
\newblock ISBN 978-0-387-72065-4.

\bibitem[Hutter and J{\"o}hnk(2004)]{hutter_continuum_2004}
K.~Hutter and K.~D. J{\"o}hnk.
\newblock \emph{Continuum methods of physical modeling: continuum mechanics,
  dimensional analysis, turbulence}.
\newblock Springer, Berlin, 2004.
\newblock ISBN 3540206191 9783540206194.

\bibitem[M\"uller et~al.(2013)M\"uller, Kummer, and
  Oberlack]{muller_highly_2013}
B.~M\"uller, F.~Kummer, and M.~Oberlack.
\newblock Highly accurate surface and volume integration on implicit domains by
  means of moment-fitting.
\newblock \emph{Int. J. Numer. Meth. Eng.}, pages 512--528, 2013.
\newblock ISSN 00295981.
\newblock \doi{10.1002/nme.4569}.

\bibitem[Sauerland and Fries(2013)]{sauerland_stable_2013}
H.~Sauerland and T.-P. Fries.
\newblock The stable {XFEM} for two-phase flows.
\newblock \emph{Computers \& Fluids}, 87:\penalty0 41--49, 2013.
\newblock ISSN 00457930.
\newblock \doi{10.1016/j.compfluid.2012.10.017}.
\newblock URL
  \url{http://linkinghub.elsevier.com/retrieve/pii/S0045793012004148}.

\bibitem[Schenk(2004)]{schenk_2004}
O.~Schenk.
\newblock Solving unsymmetric sparse systems of linear equations with
  {PARDISO}.
\newblock \emph{Future Generation Computer Systems}, 20\penalty0 (3):\penalty0
  475--487, 2004.
\newblock ISSN {0167739X}.
\newblock \doi{10.1016/j.future.2003.07.011}.

\bibitem[Schenk and G\"artner(2006)]{schenk_2006}
O.~Schenk and K.~G\"artner.
\newblock On fast factorization pivoting methods for sparse symmetric
  indefinite systems.
\newblock \emph{Electronic Transactions on Numerical Analysis}, 23:\penalty0
  158--179, 2006.

\bibitem[Schenk et~al.(2000)Schenk, G\"artner, and Fichtner]{schenk_2000}
O.~Schenk, K.~G\"artner, and W.~Fichtner.
\newblock Efficient sparse {LU} factorization with {Left-Right} looking
  strategy on shared memory multiprocessors.
\newblock \emph{{BIT} Numerical Mathematics}, 40\penalty0 (1):\penalty0
  158--176, 2000.
\newblock ISSN 0006-3835.

\bibitem[Wang and Oberlack(2011)]{wang_thermodynamic_2011}
Y.~Wang and M.~Oberlack.
\newblock A thermodynamic model of multiphase flows with moving interfaces and
  contact line.
\newblock \emph{Continuum Mechanics and Thermodynamics}, 23:\penalty0 409--433,
  May 2011.
\newblock ISSN 0935-1175, 1432-0959.

\bibitem[Zienkiewicz and Zhu(1992)]{zienkiewicz_superconvergent_1992}
O.~Zienkiewicz and J.~Zhu.
\newblock The superconvergent patch recovery ({SPR}) and adaptive finite
  element refinement.
\newblock \emph{Computer Methods in Applied Mechanics and Engineering},
  101\penalty0 (1-3):\penalty0 207--224, 1992.
\newblock ISSN 00457825.
\newblock \doi{10.1016/0045-7825(92)90023-D}.
\newblock URL
  \url{http://linkinghub.elsevier.com/retrieve/pii/004578259290023D}.

\end{thebibliography}

\end{document}